\documentclass[prd,aps,nofootinbib,superscriptaddress,floatfix,10 pt]{revtex4}%
\usepackage{amssymb}
\usepackage{amsmath,graphicx,color,epstopdf,graphics}
\usepackage{subfigure}
\usepackage{amsmath}
\usepackage{amsfonts}
\usepackage{graphicx}%
\usepackage{natbib}

\begin{document}
\preprint{ }
\title{Magneto-optical properties of topological insulator thin films with broken
inversion symmetry}
\author{Kulsoom Rahim}
\affiliation{Department of Physics, Quaid-i-Azam University, Islamabad 45320, Pakistan}
\author{Ahsan Ullah}
\affiliation{Department of Physics, Quaid-i-Azam University, Islamabad 45320, Pakistan}
\author{Muhammad Tahir}
\affiliation{Department of Physics, College of Science, University of Hafr Al Batin, Hafr Al Batin 31991, Kingdom of Saudi Arabia}
\author{Kashif Sabeeh}
\email{ksabeeh@qau.edu.pk}
\affiliation{Department of Physics, Quaid-i-Azam University, Islamabad 45320, Pakistan}
\affiliation{}

\begin{abstract}
We determine the optical response of ultrathin film topological insulators in
the presence of a quantizing external magnetic field taking into account both
hybridization between surface states, broken inversion symmetry and explicit
time reversal symmetry breaking by the magnetic field. We find that breaking
of inversion symmetry in the system, which can be due to interaction with a
substrate or electrical gating, results in Landau level crossings which lead
to additional optical transition channels that were previously forbidden. We
show that by tuning the hybridization and symmetry breaking parameters, a
transition from the normal to a topological insulator phase occurs with
measureable signatures in both the longitudinal and optical Hall conductivity.

\end{abstract}
\startpage{01}
\endpage{02}
\maketitle

\section{Introduction}

Topological insulators fall in a class of materials known as symmetry
protected topological phases. In these systems the gapless Dirac spectrum of
the surface states is protected by symmetries such as charge conservation,
time reversal and spatial inversion. Breaking of these symmetries leads to
opening of a gap in the spectrum. In this work, we consider a subclass of
these systems which are ultrathin topological insulators (TIs) with gapped
Dirac cones on their top and bottom surfaces. Thin films of topological
insulators have been experimentally fabricated for various TI materials such
as $Sb_{2}Te_{3}$ \cite{1a}. The gap in the top and bottom surface states is
usually called hybridization gap which arises as a result of coupling between
top and bottom surface states of a 3D topological insulator when its thickness
is sufficiently small, 5QLs (Quintuple Layers) and thinner \cite{1b,1c,1d}. This
gap can be tuned by varying the thickness of topological insulator
films \cite{1e}. Hence in addition to breaking of the aforementioned symmetries,
gap opening in thin films is also possible through hybridization. We
investigate the role of gap opening through symmetry breaking and
hybridization on the optical response of the system. Specifically, we study
the magneto-optical response of TI thin films with inversion symmetry
breaking. Inversion symmetry breaking occurs in thin films grown on a
substrate and can also be tuned by electrical gating \cite{1c,1f}. In both
cases, chemical potential on the two surface can be different with the result
that the Dirac points are not at the same energy on the two surfaces. In
addition to gap tuning, another advantage of thin films is that their bulk
contribution is small and allows observation of surface properties of
topological insulators \cite{1g,1h,1i}. In earlier work, it was revealed that thin
film topological insulators show interesting physics when time reversal
symmetry is broken; either by magnetic ordering or by the application of
external magnetic field \cite{1j,1k,1l,1m,1n,1o}. It was shown that through gap
tuning, the system can make a transition from the normal insulator(NI) to a TI
phase with finite dc Hall conductivity \cite{1k,1m,1o}.

The main focus of the present work is the investigation of inversion symmetry
breaking effects on the magneto-optical response of TI thin films. In addition
to inversion symmetry, time reversal symmetry can also be explicitly broken by
a magnetic field applied to our system, which we consider. Inversion symmetry
breaking generates additional gap in the spectrum of topological insulator
thin films \cite{1c}. This energy gap is not only controlled by thickness of the
films \cite{1e} and external exchange field/magnetic field \cite{1j}, but it can
also be generated through interaction with a substrate and can be tuned by
electrical gating \cite{2a}. In this paper, we show that in the presence of
inversion symmetry breaking, a new feature of the Landau level spectrum is
Landau level crossings which lead to new optical transition channels that can
be observed in the optical response. These channels were previously forbidden
in the presence of inversion symmetry. Further, we determine the dc and the
optical conductivity in the quantum Hall regime and show the presence of Hall
steps and plateaus even in the ac regime. We also show that by tuning the
hybridization and symmetry breaking parameters, a transition from the normal
to a topological insulator phase occurs with measurable signatures in the
magneto-optical response.

The paper is organized as follows: In Sec. II, we present our model system.
Sec. III is devoted to the calculation of the optical conductivity tensor for
our system. Results for optical Hall conductivity in the quantum Hall regime
are presented in Sec. IV with conclusions and summary of results in Sec. V.

\section{Model}

Our system is a topological insulator thin film, thin enough for hybridization
of surface states on the top and bottom, and broken inversion symmetry which
can be either due to gating or substrate. In order to highlight the effects of
Time Reversal (TR) symmetry breaking we include a term in the Hamiltonian
which can arise due to magnetic ordering in the presence of magnetic dopants
in the proximity of the top surface which are exchange coupled to the
electronic spins. This term has been included here to illustrate the effects
of TR symmetry breaking on the energy spectrum. The effective low energy
Hamiltonian is \cite{1c,1j,1f}%

\begin{equation}
H=\hbar v_{f}\tau_{z}\otimes(\sigma_{x}k_{y}-\sigma_{y}k_{x})+\Delta_{h}%
\tau_{x}\otimes I+\Delta_{ib}\tau_{z}\otimes I+\Delta_{z}I \otimes\sigma_{z},
\label{1}%
\end{equation}

with the basis: $|t\uparrow\rangle,|t\downarrow\rangle,|b\uparrow\rangle$ and
$|b\downarrow\rangle.$ Here $t,b$ denote the top and bottom surface states and
$\uparrow,\downarrow$ represent the spin up and down states. $v_{f}$ is the
Fermi velocity, and $I$ is the identity matrix, $\sigma_{i}(i=x,y,z)$ and
$\tau_{j}(j=x,y,z)$ are Pauli matrices acting on spin space and opposite
surface space (surface pseudospin). $\Delta_{h}$ represents the hybridization
between the two surface states. For large thickness $\Delta_{h}\approx0$,
hybridization of top and bottom surface states can be neglected, however when
thickness is sufficiently reduced $\Delta_{h}$ generates finite gap in the
Dirac spectrum. $\Delta_{ib}$ is the inversion symmetry breaking term between
the two surfaces, it can result from interaction between the TI thin film and
the substrate or by an electric field applied perpendicular to the surface of
the thin film \cite{1c}. $\Delta_{z}$ can be the exchange field along the z-axis
introduced by possible ferromagnetic ordering of the magnetic impurities. The
energy spectrum of the above Hamiltonian is given by%
\begin{equation}
\varepsilon_{\pm}^{\alpha}(k)=(-1)^{\alpha}\sqrt{\hbar^{2}v_{f}^{2}%
k^{2}+\Delta_{h}^{2}+\Delta_{ib}^{2}+\Delta_{z}^{2}\pm2\sqrt{\hbar^{2}%
v_{f}^{2}k^{2}\Delta_{ib}^{2}+\Delta_{ib}^{2}\Delta_{z}^{2}+\Delta_{z}%
^{2}\Delta_{h}^{2}}},
\end{equation}
where $\alpha=1$ represents the states in valence band and $\alpha=0$
represents the states in conduction band. $\pm$ \ correspond to upper and
lower surface. Fig. (1) shows the band structure for our system at different
values of $\Delta_{h},\Delta_{ib}$ and $\Delta_{z}$. For $\Delta_{h}%
=\Delta_{ib}=\Delta_{z}=0$, both top and bottom surface states are gapless and
degenerate. For a thin TI without any source of TR and I symmetry breaking,
$(\Delta_{h}\neq0;\Delta_{ib}=\Delta_{z}=0),$ the bands are degenerate
separated by insulating gap $\Delta_{h}$. For the case of inversion asymmetry
and finite hybridization with the system preserving time reversal symmetry
$(\Delta_{z}=0)$, a Rashba-like splitting in the band structure occurs; Fig.
1(c). The bands are degenerate at $k=0$ as $\varepsilon_{+}^{0}%
(k=0)=\varepsilon_{-}^{0}(k=0)$ in conduction band and $\varepsilon_{+}%
^{1}(k=0)=\varepsilon_{-}^{1}(k=0)$ in valence band for all values of
$\Delta_{h}$ and $\Delta_{ib}$ while for $k\neq0$ the bands are not
degenerate. This $k=0$ degeneracy is lifted by the introduction of time
reversal symmetry breaking term $\Delta_{z}$, with the result that the
degeneracy does not exist for any value of $k$. Therefore, the band structure
represents Normal Insulator (NI) regime for small value of $\Delta_{z}$, as
the value of $\Delta_{z}$ is increased the lower gap decreases. At a
particular value of $\Delta_{z}$ the gap closes. This gapless point represents
the phase transition point. As $\Delta_{z}$ is further increased the
transition from NI to TI phase takes place and the gap reopens, (see
Figs.(1e)-(1f)). The red lines (the blue lines) represent dispersion of upper
(lower) surface.\newline Now we consider the effect of Landau quantization on
the system by the application of an external magnetic field along $z$-axis
directed perpendicular to the surface of TI thin film aligned in the xy-plane.
The magnetic field explicitly breaks TR symmetry. In the rest of the paper we
will not consider the effect of magnetic ordering. The Hamiltonian of our
system takes the form \cite{1m,1o},%
\[
\hat{H}_{\sigma\tau}=\hbar v_{f}\tau_{z}\otimes\left[  \sigma_{x}\mathbf{\pi
}_{y}-\sigma_{y}\mathbf{\pi}_{x}\right]  +\Delta_{h}\tau_{x}\otimes
I+\Delta_{ib}\tau_{z}\otimes I+\Delta_{z}I\otimes\sigma_{z},
\]
$\mathbf{\pi=k+}e\mathbf{A/}\hbar$ is the two dimensional canonical momentum
with vector potential $\mathbf{A}$. Here $\Delta_{z}=g%
\mu
_{B}B/2$ is the Zeeman energy associated with the applied magnetic field
$B=B\hat{z}$, with $g$ effective Lande factor, $\mu_{B}$ is the Bohr magneton.
We choose the Landau gauge for the vector potential $A=(0,xB,0).$ Since
$p_{x}$ and $x$ do not commute, it is convenient to write the Hamiltonian in
terms of dimensionless operators%
\[
\hat{H}_{\sigma\tau}=\frac{\hbar v_{f}}{\sqrt{2}l_{B}}\tau_{z}\otimes\left[
\sigma_{x}l_{B}\hat{P}+\sigma_{y}\frac{\hat{Q}}{l_{B}}\right]  +\Delta_{h}%
\tau_{x}\otimes I+\Delta_{ib}\tau_{z}\otimes I+\Delta_{z}I\otimes\sigma_{z},
\]
where $l_{B}=\sqrt{c/eB}$ is the magnetic length. $\hat{Q}=-l_{B}^{2}p_{x}$
and $\hat{P}=p_{y}+\frac{eB}{\hbar}x$ such that $[\hat{Q},\hat{P}]=i\hbar$.
Employing the ladder operators $a=1/\sqrt{2}l_{B}(\hat{Q}+il_{B}^{2}\hat{P})$
and $a^{\dagger}=1/\sqrt{2}l_{B}(\hat{Q}-il_{B}^{2}\hat{P}),$ we may express
the Hamiltonian as:%
\begin{equation}
H=\frac{i\hbar\omega_{B}}{\sqrt{2}}\tau_{z}\otimes(\sigma^{+}a-\sigma
^{-}a^{\dagger})+\Delta_{h}\tau_{x}\otimes I+\Delta_{ib}\tau_{z}\otimes
I+\Delta_{z}I\otimes\sigma_{z} \label{3}%
\end{equation}
where $\omega_{B}=v_{F}/l_{B}$, which plays a role analogous to the cyclotron
frequency in the LL spectrum of a regular 2DEG. We can write single particle
eigenstates in the following form%
\begin{equation}
\left\vert n\alpha s\right\rangle =u_{nT\uparrow}^{\alpha s}\left\vert
n-1,T,\uparrow\right\rangle +u_{nT\downarrow}^{\alpha s}\left\vert
n,T,\downarrow\right\rangle +u_{nB\uparrow}^{\alpha s}\left\vert
n-1,B,\uparrow\right\rangle +u_{nB\downarrow}^{\alpha s}\left\vert
n,B,\downarrow\right\rangle . \label{4}%
\end{equation}
Here $\left\vert n,T(B),\uparrow(\downarrow)\right\rangle $ is the nth LL
eigenstates on the top (bottom) surface with spin up (down), $\alpha=0,1$ and
$s=\pm$ label the four eigenstates of Eq. (\ref{3}), corresponding to each LL
index $n=0,...,\infty$, and $u_{n}^{\alpha s}$ are the corresponding complex
four component spinor wave functions. Thus the Hamiltonian Eq. (\ref{3}) can be
written in a $4\times4$ matrix%
\begin{equation}
H=%
\begin{pmatrix}
\Delta_{z}+\Delta_{ib} & -i\hbar\omega_{B}\sqrt{2n} & \Delta_{h} & 0\\
i\hbar\omega_{B}\sqrt{2n} & -\Delta_{z}+\Delta_{ib} & 0 & \Delta_{h}\\
\Delta_{h} & 0 & \Delta_{z}-\Delta_{ib} & i\hbar\omega_{B}\sqrt{2n}\\
0 & \Delta_{h} & -i\hbar\omega_{B}\sqrt{2n} & -(\Delta_{z}+\Delta_{ib})
\end{pmatrix}
., \label{5}%
\end{equation}
Diagonalizing the hamiltonian in Eq. (\ref{5}), we find the following Landau
level spectrum,%
\begin{equation}
\epsilon_{n\alpha s}=(-1)^{\alpha}\sqrt{\Delta_{ib}^{2}+\Delta_{h}^{2}%
+\Delta_{z}^{2}+2n\hbar^{2}\omega_{B}^{2}+2s\sqrt{\Delta_{ib}^{2}\Delta
_{z}^{2}+\Delta_{h}^{2}\Delta_{z}^{2}+2n\Delta_{ib}^{2}\hbar^{2}\omega_{B}%
^{2}}}. \label{6}%
\end{equation}

The Landau level energy spectrum is shown in Fig. (2). An important feature of
the spectrum is that its electron-hole symmetric for $n\neq0.$ The LL spectrum
consists of two sets: $(i)\epsilon_{n0s}$ and $(ii)\epsilon_{n1s}$ where
$\epsilon_{n1s}$ represents spectrum for occupied states below $\mu=0$ and
$\epsilon_{n0s}$ represent the unoccupied states above $\mu=0$. For both sets
of occupied and unoccupied Landau levels each Landau level splits in a doublet
for $s=\pm1$. This splitting results from time reversal symmetry breaking term
$(\Delta_{z}),$ broken inversion symmetry and the hybridization $(\Delta_{h})$
in the Hamiltonian. There are two situations where splitting can vanish, if
$(i)$ the system has inversion symmetry $(\Delta_{ib}=0)$ and broken TR
symmetry along with no hybridization $(\Delta_{h}=0)$ $(ii)$ system has both
TR symmetry$(\Delta_{z}=0)$ and inversion symmetry$(\Delta_{ib}=0)$ with no
constraint on the hybridization (for zero hybridization with both TR and
inversion symmetry thin film TI will behave like gapless Dirac material with
only one $n=0$ partially filled LL at $\mu=0$). Note that the $n=0$ LL, only
splits when either $\Delta_{h}$ or $\Delta_{ib}$ is nonzero. A novel feature
of LLs in the presence of broken inversion symmetry is their crossings within
each set, between $nth$ and $(n+1)th$ levels with opposite $s$ values, at
certain values of magnetic field. There is no crossing of $n=0$ LL in our
system at any value of magnetic field. We consider terms linear in $k$.
However, for large values of $k$ hybridization terms must be redefined as
$\frac{\Delta}{2}-Bk^{2}$. This results in crossing of the $n=0$ Landau
levels. This crossing becomes an anti-crossing in the presence of inversion
symmetry breaking \cite{2b}

The $n=0$ LLs behaves differently compared to $n\neq0$ LLs. For $n=0$ Eq.
(\ref{4}) shows that the electrons are fully spin-polarized in these levels;
only spin down levels are occupied. Hence $n=0$ levels are split into two
sublevels with spin down unlike other levels which are split into four
sublevels, two with spin up and two with spin down. The corresponding $n=0$
wavefunctions (un-normalized) are%
\[
u_{0}^{\alpha s}=\{0,s(-1)^{\alpha}\frac{(-\Delta_{ib}+s\sqrt{\Delta_{ib}%
^{2}+\Delta_{h}^{2}})}{\Delta_{h}},0,1\}.
\]
The corresponding LL energies are
\begin{equation}
\epsilon_{0\alpha s}=(-1)^{\alpha}\left\vert \Delta_{z}+s\sqrt{\Delta_{ib}%
^{2}+\Delta_{h}^{2}}\right\vert .\nonumber
\end{equation}
Explicitly
\begin{align}
u_{0}^{0-1}  &  =\{0,\frac{\Delta_{ib}+\sqrt{\Delta_{ib}^{2}+\Delta_{h}^{2}}%
}{\Delta_{h}},0,1\},\label{7}\\
u_{0}^{11}  &  =\{0,\frac{\Delta_{ib}-\sqrt{\Delta_{ib}^{2}+\Delta_{h}^{2}}%
}{\Delta_{h}},0,1\},\nonumber
\end{align}
for $\Delta_{z}<\sqrt{\Delta_{ib}^{2}+\Delta_{h}^{2}}.$%

\begin{equation}
\epsilon_{00-1}=\left\vert \Delta_{z}-\sqrt{\Delta_{ib}^{2}+\Delta_{h}^{2}%
}\right\vert ,\text{ \ \ \ \ \ }\epsilon_{011}=-\left\vert \Delta_{z}%
+\sqrt{\Delta_{ib}^{2}+\Delta_{h}^{2}}\right\vert . \label{8}%
\end{equation}
For $\Delta_{z}>\sqrt{\Delta_{ib}^{2}+\Delta_{h}^{2}},$ $u_{0}^{0-1}$ is
replaced by $u_{0}^{1-1}:$
\begin{equation}
u_{0}^{1-1}=\{0,\frac{\Delta_{ib}+\sqrt{\Delta_{ib}^{2}+\Delta_{h}^{2}}%
}{\Delta_{h}},0,1\},\nonumber
\end{equation}
with energy given by%
\[
\epsilon_{01-1}=-\left\vert \Delta_{z}-\sqrt{\Delta_{ib}^{2}+\Delta_{h}^{2}%
}\right\vert .
\]
In the NI phase the states are $u_{0}^{0-1}$ and $u_{0}^{11}$, and in TI phase
the states are $u_{0}^{1-1}$ and $u_{0}^{11}.$ Thus one of the $n=0$
electron-like sublevel becomes a hole-like sublevel which can be seen in Fig.
(3) for density of states and in the Landau level spectrum in Fig. (2). This
change in character of the zeroth LL manifests particle-hole symmetry breaking
in the system which results in jump in Hall conductivity from $0$ to a finite
value at chemical potential $\mu=0$. This signals a transition from the NI
phase to the TI phase.

\subsection{Density of states}

The Green function associated with our Hamiltonian is
\[
G(\omega,n,\alpha,s)=%
{\displaystyle\sum\limits_{\alpha,s}}
\frac{1}{\omega-(-1)^{\alpha}\sqrt{\Delta_{ib}^{2}+\Delta_{h}^{2}+\Delta
_{z}^{2}+2n\hbar^{2}\omega_{B}^{2}+2s\sqrt{\Delta_{ib}^{2}\Delta_{z}%
^{2}+\Delta_{h}^{2}\Delta_{z}^{2}+2n\Delta_{ib}^{2}\hbar^{2}\omega_{B}^{2}}%
}+i\eta}.
\]

From which we can compute density of states as%
\begin{align*}
D(\omega)  &  =\frac{-1}{2\pi l_{B}^{2}}%
{\displaystyle\sum\limits_{n}}
ImG(\omega,n,\alpha,s),\\
D(\omega)  &  =\frac{-1}{\pi}\frac{1}{2\pi l_{B}^{2}}(%
{\displaystyle\sum\limits_{n\neq0}}
Im G(\omega,n,\alpha,s)+Im G(\omega,0,\alpha,s)).
\end{align*}

The plots for density of states are shown in Fig. (3). $(NI)$ represents the
normal insulator phase in which the LL spectrum has perfect particle-hole
symmetry. At $\Delta_{z}>\sqrt{\Delta_{ib}^{2}+\Delta_{h}^{2}},$ both $n=0$
levels are filled as they shift to the valence band, thus breaking
particle-hole symmetry for LL spectrum across $\epsilon=0$. This shift in the
Landau level is associated with transition from (NI) phase to the TI phase. We
notice two interesting features of the density of states: (i) At certain
values of $\omega$ two adjacent peaks are closely spaced such that they tend
to merge in a single peak. (ii) Some peaks have higher weight than all other
peaks. These two distinct features are attributed to crossing (or near
crossing) of two Landau levels at certain values of magnetic field, see Fig.
(2). The parameters chosen in these plots are $\Delta_{ib}=0.006eV$, $\Delta
_{h}=0.004eV$ and $\hbar v_{f}^{2}eB=1.6\times10^{-4}B$.

\section{Magneto-optical conductivity}

To determine the magneto-optical conductivity tensor we need eigenfunctions of
the Hamiltonian given in Eq. (5). The explicit form of the eigenfunctions
$u_{n}^{\alpha s}$ in Eq. (4) is
\begin{align}
u_{n}^{\alpha s}  &  =(u_{nT\uparrow}^{\alpha s},u_{nT\downarrow}^{\alpha
s},u_{nB\uparrow}^{\alpha s},u_{nB\downarrow}^{\alpha s}),\label{11}\\
u_{nT\uparrow}^{\alpha s}  &  =\dfrac{-i}{d_{1}N}(N_{1}+sN_{2}(\Delta
_{ib}+\Delta_{z})+\epsilon_{n\alpha s}(\Delta_{ib}\Delta_{z}+sN_{2})),\\
u_{nT\downarrow}^{\alpha s}  &  =\dfrac{1}{d_{2}N}(\Delta_{ib}^{2}%
+sN_{2}+\epsilon_{n\alpha s}\Delta_{ib}),\\
u_{nB\uparrow}^{\alpha s}  &  =\dfrac{-1}{d_{3}N}(N_{3}-\Delta_{h}%
(-\Delta_{ib}-\Delta_{z}-\epsilon_{n\alpha s})(\Delta_{ib}-\Delta_{z}%
-\epsilon_{n\alpha s})),\\
u_{nB\downarrow}^{\alpha s}  &  =\dfrac{1}{N}.
\end{align}
$N_{1}$,$N_{2}$,$N_{3}$,$d_{1}$,$d_{2}$,$d_{3}$ are given by
\begin{align}
d_{1}  &  =\sqrt{2n}\Delta{_{h}}\omega{_{B}}(\Delta_{ib}-\Delta_{z}%
)\label{12}\\
d_{2}  &  =\Delta{_{h}}(\Delta_{ib}-\Delta_{z})\\
d_{3}  &  =2i\sqrt{2n}\Delta_{h}\omega_{B}(\Delta_{ib}-\Delta_{z})\\
N_{1}  &  =\Delta_{ib}^{2}\Delta_{z}+\Delta_{h}^{2}\Delta_{z}+\Delta
_{ib}\Delta_{z}^{2}+2n\omega_{B}^{2}\Delta_{ib}\\
N_{2}  &  =\sqrt{\Delta_{ib}^{2}\Delta_{z}^{2}+\Delta_{h}^{2}\Delta_{z}%
^{2}+2n\omega_{B}^{2}\Delta_{ib}^{2}}\\
N_{3}  &  =\Delta_{h}(\Delta_{h}^{2}+2n\omega_{B}^{2}),
\end{align}
and $N$ is the normalization constant. Given the above eigenfunctions we can
now evaluate the magneto-optical conductivity tensor within linear response
regime using Kubo formula \cite{2c}.
\begin{equation}
\sigma_{\alpha\beta}(\omega)=\dfrac{i\hbar}{2\pi l_{B}^{2}}\underset{n\alpha
s\neq n^{\prime}\alpha^{\prime}s^{^{\prime}}}{\sum}\dfrac{f(\varepsilon
_{n\alpha s})-f(\varepsilon_{n^{\prime}\alpha^{\prime}s^{\prime}})}%
{\epsilon_{n^{\prime}\alpha^{\prime}s^{\prime}}-\epsilon_{n\alpha s}}%
\dfrac{\left\langle n\alpha s\right\vert \hat{\jmath}_{\alpha}\left\vert
n^{\prime}\alpha^{\prime}s^{\prime}\right\rangle \left\langle n^{\prime}%
\alpha^{\prime}s^{\prime}\right\vert \hat{\jmath}_{\beta}\left\vert n\alpha
s\right\rangle }{\hbar\omega-\epsilon_{n^{\prime}\alpha^{\prime}s^{\prime}%
}+\epsilon_{n\alpha s}+i\hbar/(2\tau)} \label{13}%
\end{equation}
where $\hat{\jmath}_{\alpha}=\frac{e}{\hbar}\dfrac{\partial H}{\partial
k_{\alpha}}$ and $f(\varepsilon_{n\alpha s})=\frac{1}{1+exp[\beta
(\epsilon_{n\alpha s}-\mu)]}$ is the Fermi distribution function with
$\beta=1/k_{B}T$ and $\mu$ is the chemical potential. We note that transitions
between occupied Landau levels are Pauli blocked. So the only allowed
transitions will be from the occupied LLs in valence band to unoccupied LLs in
conduction band (i.e. across chemical potential $\mu=0$). After evaluating the
matrix elements we have found that the selection rules for allowed transitions
is $n^{\prime}=n\pm1$. The absorptive part of the conductivity for $n=0$
Landau level above and below $\mu=0$ is
\begin{align}
\binom{\operatorname{Re}\sigma_{xx}(\omega)/\sigma_{0}}{Im\sigma_{xy}%
(\omega)/\sigma_{0}}  &  =\hbar ev_{f}^{2}B\underset{s}{[\sum}\dfrac
{(f(\epsilon_{11s})-f(\epsilon_{00-1}))N(1,0,s,1,1)\times\eta}{(\epsilon
_{00-1}-\epsilon_{11s})((\hbar\omega+\epsilon_{11s}-\epsilon_{00-1})^{2}%
+\eta^{2})}\nonumber\\
&  \pm\underset{s}{\sum}\dfrac{(f(\epsilon_{011})-f(\epsilon_{10s}%
))M(0,0,-1,1,s)\times\eta}{(\epsilon_{10s}-\epsilon_{011})((\hbar
\omega+\epsilon_{011}-\epsilon_{10s})^{2}+\eta^{2})}\nonumber\\
\pm &  \underset{n=2,ss^{\prime}\alpha\neq\alpha^{\prime}}{\sum}%
\dfrac{(f(\epsilon_{n\alpha s})-f(\epsilon_{n+1\alpha^{\prime}s^{\prime}%
}))M(n,s,s^{^{\prime}}\alpha,\alpha^{\prime})\times\eta}{(\epsilon
_{n+1\alpha^{\prime}s^{\prime}}-\epsilon_{n\alpha s})((\hbar\omega
+\epsilon_{n\alpha s}-\epsilon_{n+1\alpha^{\prime}s^{\prime}})^{2}+\eta^{2}%
)}\nonumber\\
&  +\underset{n=2,s,s^{\prime}\alpha\neq\alpha^{\prime}}{\sum}\dfrac
{(f(\epsilon_{n\alpha s})-f(\epsilon_{n-1\alpha^{\prime}s^{\prime}%
}))N(n,s,s^{\prime},\alpha,\alpha^{\prime})\times\eta}{(\epsilon_{n-1\alpha
^{\prime}s^{\prime}}-\epsilon_{n\alpha s})((\hbar\omega+\epsilon_{n\alpha
s}-\epsilon_{n-1\alpha^{\prime}s^{\prime}})^{2}+\eta^{2})}], \label{14}%
\end{align}
and for the case when both $n=0$ LLs are hole-like Landau levels, the
conductivity is%
\begin{align}
\binom{\operatorname{Re}\sigma_{xx}(\omega)/\sigma_{0}}{Im\sigma_{xy}%
(\omega)/\sigma_{0}}  &  =\hbar ev_{f}^{2}B\underset{s}{[\sum}\dfrac
{(f(\epsilon_{01-1})-f(\epsilon_{10s}))N(1,0,s,1,1)\times\eta}{(\epsilon
_{10s}-\epsilon_{01-1})((\hbar\omega+\epsilon_{01-1}-\epsilon_{10s})^{2}%
+\eta^{2})}\nonumber\\
&  \pm\underset{s}{\sum}\dfrac{(f(\epsilon_{011})-f(\epsilon_{10s}%
))M(0,1,-1,0,s)\times\eta}{(\epsilon_{10s}-\epsilon_{011})((\hbar
\omega+\epsilon_{011}-\epsilon_{10s})^{2}+\eta^{2})}\nonumber\\
\pm &  \underset{n=1,\alpha s\neq\alpha^{\prime}s^{^{\prime}}}{\sum}%
\dfrac{(f(\epsilon_{n\alpha s})-f(\epsilon_{n+1\alpha^{\prime}s^{\prime}%
}))M(n,\alpha,s,\alpha^{\prime},s^{^{\prime}})\times\eta}{(\epsilon
_{n+1\alpha^{\prime}s^{\prime}}-\epsilon_{n\alpha s})((\hbar\omega
+\epsilon_{n\alpha s}-\epsilon_{n+1\alpha^{\prime}s^{\prime}})^{2}+\eta^{2}%
)}\nonumber\\
&  +\underset{n=2,s,s^{\prime}\alpha\neq\alpha^{\prime}}{\sum}\dfrac
{(f(\epsilon_{n\alpha s})-f(\epsilon_{n-1\alpha^{\prime}s^{\prime}})%
)N(n,\alpha,s,\alpha^{\prime},s^{^{\prime}})\times\eta}{(\epsilon
_{n-1\alpha^{\prime}s^{\prime}}-\epsilon_{n\alpha s})((\hbar\omega
+\epsilon_{n\alpha s}-\epsilon_{n-1\alpha^{\prime}s^{\prime}})^{2}+\eta^{2}%
)}]. \label{15}%
\end{align}
Here $\sigma_{0}=e^{2}/h,$ $\eta=\hbar/2\tau$ is the scattering rate related
to broadening of the Landau levels and%
\begin{equation}
M(n,\alpha,s,\alpha^{\prime},s^{^{\prime}})=[(u_{nT\downarrow}^{\alpha
s})^{\ast}u_{n+1T\uparrow}^{\alpha^{\prime}s^{\prime}}-(u_{nB\downarrow
}^{\alpha s})^{\ast}u_{n+1B\uparrow}^{\alpha^{\prime}s^{\prime}}]\times
\lbrack u_{nT\downarrow}^{\alpha s}(u_{n+1T\uparrow}^{\alpha^{\prime}%
s^{\prime}})^{\ast}-u_{nB\downarrow}^{\alpha s}(u_{n+1B\uparrow}%
^{\alpha^{\prime}s^{\prime}})^{\ast}],
\end{equation}
and%
\begin{equation}
N(n,\alpha,s,\alpha^{\prime},s^{^{\prime}})=[(u_{nB\uparrow}^{\alpha s}%
)^{\ast}u_{n-1B\downarrow}^{\alpha^{\prime}s^{\prime}}-(u_{nT\uparrow}^{\alpha
s})^{\ast}u_{n-1T\downarrow}^{\alpha^{\prime}s^{\prime}}]\times\lbrack
u_{nB\uparrow}^{\alpha s}(u_{n-1B\uparrow}^{\alpha^{\prime}s^{\prime}})^{\ast
}-u_{nT\uparrow}^{\alpha s}(u_{n-1T\downarrow}^{\alpha^{\prime}s^{\prime}%
})^{\ast}].
\end{equation}

Where $\ast$ denotes complex conjugation. First let us consider the
conductivity at zero temperature. The features of the spectrum to bear in mind
are: All $n\neq0$ levels are split in a doublet $(s=\pm)$ for finite value of
Zeeman energy and hybridization. The inversion symmetry breaking term
$\Delta_{ib}$ results in crossing of Landau levels at different values of
magnetic field strength. Therefore there are not only the allowed transitions
between LLs with same $s$ but transitions can also occur between LLs with
different $s.$ These transitions are not allowed in the presence of inversion
symmetry\cite{1o}. Hence inversion symmetry breaking opens optical transition
channels which were previously forbidden.

Fig. (4) shows results for absorptive peaks from Eq. (\ref{14}) and Eq. (\ref{15}).
All the absorptive peaks result from the transitions between LLs across
$\mu=0$. In Fig. (4(a,b)), in the NI phase, the first peak corresponds to
transition from $n=0$ and $n=1$ Landau level with $\omega=\epsilon
_{00-1}-\epsilon_{11-1}.$ This is a single transition peak and is close to the
2nd absorption peak between the LL, $\omega=\epsilon_{10-1}-\epsilon_{011}$.
Next two peaks are small and are also contributed by the Landau level
transition between $n=0$ and $n=1$. The set of transitions involving $n=0$ and
$n=1$ in NI phase are $\epsilon_{11s}\rightarrow\epsilon_{00-1}$ and
$\epsilon_{011}\rightarrow\epsilon_{10s}$. As the magnetic field is increased
which increases Zeeman energy such that at $\Delta_{z}=\sqrt{\Delta_{ib}%
^{2}+\Delta_{h}^{2}}$ , the LL $\epsilon_{00-1}$ becomes partially filled. At
this stage one of the peaks corresponding to $n=0$ and $n=1$ transition
disappears involving $n=0$ are $\epsilon_{011}\rightarrow\epsilon_{10s}$.
Increasing Zeeman energy further by increasing the magnetic field such that it
is greater then $4.25T$ the $n=0$ doublet becomes hole like doublet and LL
$\epsilon_{00-1}$ changes to $\epsilon_{01-1}$. The allowed transitions for
$n=0$ are $\epsilon_{011}\rightarrow\epsilon_{10s}$ and $\epsilon
_{01-1}\rightarrow\epsilon_{10s}$ represented by peaks at $\omega
=\epsilon_{10s}-\epsilon_{01-1}$ and $\omega=\epsilon_{10s}-\epsilon_{011}$.
The transitions $\epsilon_{11s}\rightarrow\epsilon_{00-1}$ in NI is replaced
by $\epsilon_{01-1}\rightarrow\epsilon_{10s}$ in the TI phase. The absorption
peaks for $n\neq0$ shift to higher energy in TI phase because at high magnetic
field the gap between LL increases.

To understand the behavior of $Im\sigma_{xy}$ we must keep in mind the minus
sign between the two terms in Eqs. (\ref{14}), (\ref{15}). The first two peaks
result from transition between $n=0$ and $n=1$ levels, one having positive
amplitude and the other having negative amplitude. The transition peaks
involving $n\neq0$ transitions have decreased in height which is due to the
negative sign. For example, the transitions from $n=2$ to $n=1$ Landau level
and $n=1$ to $n=2$ have same denominator and the mismatch between numerators
of the two transitions results in a net negative amplitude. Fig. (5a) shows
the imaginary part transverse conductivity for absorption peaks due to photon
absorption in the NI phase and Fig. (5b) shows the imaginary part transverse
conductivity in the TI phase.

The behavior of absorption peaks at finite chemical potential is shown in Fig.
(6) and Fig. (7). For finite value of $\mu$ in the conduction band, in addition to
filled LLs in valence band there are also filled LLs in conduction band
resulting in allowed transition within same band(intraband transitions). Here
interband and intraband transitions are defined with respect to the position
of the chemical potential. These intraband absorption peaks shift to lower
energy and do not split. The allowed transition within same bands (intraband
transitions) have greater probability as compared to interband transitions.

Fig. (9) and Fig. (10)\textbf{ }illustrate the allowed transitions between LLs at
different values of magnetic field. The blue lines are for $s=-1$ LLs and
green lines are for $s=1$ LLs. Except for $n=0$ all the other Landau levels
have perfect particle-hole symmetry. The chemical potential is represented by
thick black line. The vertical arrows show the allowed inter band transitions.
The shift of chemical potential from $\mu=0$ to some finite value results in
additional intraband transitions.

\section{Optical Hall Conductivity}

To calculate Hall conductivity we use wavefunctions from Eq. (4) in the Kubo
formula given in Eq. (20) and obtain%
\begin{align}
\sigma_{\alpha\beta}(\omega)  &  =\hbar ev_{f}^{2}B\underset{n\alpha s\neq
n^{\prime}\alpha^{\prime}s^{^{\prime}}}{\sum}\dfrac{(f(\varepsilon_{n\alpha
s})-f(\varepsilon_{n+1\alpha^{\prime}s^{\prime}}))}{(\epsilon_{n+1\alpha
^{\prime}s^{\prime}}-\epsilon_{n\alpha s})}M(n,\alpha,s,\alpha^{\prime
},s^{^{\prime}})\\
&  [\dfrac{1}{\hbar\omega-\epsilon_{n\alpha s}+\epsilon_{n+1\alpha^{\prime
}s^{\prime}}+i\hbar/(2\tau)}-\dfrac{1}{\hbar\omega-\epsilon_{n+1\alpha
^{\prime}s^{\prime}}+\epsilon_{n\alpha s}+i\hbar/(2\tau)}]
\end{align}

The effects of broken inversion symmetry that are reflected in the LL spectrum
and crossing of LLs are also revealed in the dc\ and optical Hall
conductivity, which we now discuss. In Figs. (11,12,13) plots with blue color
show the results for dc Hall conductivity ($\sigma_{xy}(\omega=0)$) with
$B=2$T (perfect particle hole symmetry across $\mu=0$), $B=4.27$T (one $n=0$
LL is partially filled at $\mu=0$) and $B=5.8$T (one extra filled LL for
negative value of $\mu$) at temperature 1K respectively. The results are
presented as a function of the chemical potential $\mu$. The LL spectrum is
also shown to emphasize the unusual behavior of steps and plateaus at specific
magnetic fields. These steps and plateaus show clear deviation from previous
results, \cite{1n}, which were in the presence of inversion symmetry. For
$n\geq1$ the plateau widths and step heights are symmetrical for both negative
and positive values of $\mu,$ reflecting particle-hole symmetry in the system.
However, widths and heights are not symmetrical within the same band because
of LL crossings. The contribution of $n=0$ LLs to Hall conductivity pleateus
shows interesting behavior. For chemical potential fixed at $\mu=0,$ the
conductivity jumps from $0$ for $\Delta_{z}<\sqrt{\Delta_{ib}^{2}+\Delta
_{h}^{2}}$ to a finite value for $\Delta_{z}>\sqrt{\Delta_{ib}^{2}+\Delta
_{h}^{2}}$. This is an indication of phase transition from NI phase to TI
phase as the magnetic field is increased. For $\Delta_{z}<\sqrt{\Delta
_{ib}^{2}+\Delta_{h}^{2}}$ the $n=0$ doublet has one electron like
($u_{0}^{0-1}$) LL and one hole like ($u_{0}^{11}$). When magnetic field is
increased such that $\Delta_{z}>\sqrt{\Delta_{ib}^{2}+\Delta_{h}^{2}}$ both
$n=0$ LLs become hole like, thus increasing the Hall conductivity. For $n=0$
LLs the Hall conductivity jump depends on magnetic field according to
$[sgn(\Delta_{z}-\sqrt{\Delta_{ib}^{2}+\Delta_{h}^{2}})+1]$. In Figs.
(10,11,12) red plots represent optical Hall conductivity. The steps and
plateaus are robust for low values of $n=(0,1)$ but as the value of $n$
increases, they no longer remain robust especially for $n>\pm2.$ We note that
as the value of magnetic field is increased the number of robust steps also
increases, which is expected. The steps structure is symmetric across $\mu=0$
except for steps involving $n=0$ LLs. The steps corresponding to low value of
$n$ are however robust unless frequency $\omega$ is close to a resonance. In
the case of static hall conductivity these steps are always robust.\newline

Next we examine the robustness of step-like structure in the optical Hall
conductivity as function of disorder strength in both the phases of the
system. The degree of disorder can be characterized by the scattering rate
parameter $\eta$ \cite{2d} We present the results of our calculation for TI
phase in Fig. (13). We can see that for dc hall conductivity the step-like
structure remains for fairly large values of $\eta$. The step like structure
for large $|n|$ begin to dminish when $\eta$ is increased. However the step
corresponding to $n=0$ LLs always remains rebust. For ac Hall conductivity the
step-like structure is less robust against increase in $\eta$. For $\eta
\simeq\omega$ the plateaus are nearly washed out for large $|n|$, while the
$n=0$ is again robust. For the case of NI phase a similar behavior is observed
; see Fig. (14).

In Fig. (15) and Fig. (16) we show real part of $\sigma_{xy}(\omega)$ as a
function of $\omega,$ we find that real part of $\sigma_{xy}(\omega)$ exhibits
sharp cyclotron resonance peaks at transition energies for transition
involving $n=0$ LL. These resonance peaks change sign near each allowed
transition frequency. As for experimental realization, since the Faraday
rotation angle is directly proportional to optical Hall conductivity, it is
possible to observe the steps in optical Hall conductivity that are predicted
here by performing Faraday rotation measurements \cite{2d,2e,2f}.

\section{Conclusions}

To conclude, we have determined the LL spectrum, density of states and the
magneto-optical conductivity tensor with in linear response regime for thin
film topological insulators with finite hybridization between the surface
states. We find that breaking of time reversal and inversion symmetry can have
profound effects on the optical response. We have shown that inversion
symmetry breaking, in addition to time reversal symmetry breaking and
hybridization, significantly affects the spectrum, transition channels and
magneto-optical response of the system. In the inversion symmetry broken TI
thin films, we have found the following: The system can exist in Normal
Insulating (NI) and Topological Insulating (TI) phases. The phase transition
between these phases can be controlled by the degree of hybridization as well
as by breaking symmetries: time reversal/inversion symmetry or both. The LL
spectrum exhibits level crossings which were not present in the inversion
symmetric system. New optical transition channels have been found which were
previously forbidden. We show that there are observable signatures of the
phase transition from NI to TI phase in both the longitudinal and optical Hall conductivity.

\section{Acknowledgement}

Kashif Sabeeh would like to acknowledge the support of the Higher Education
Commission (HEC) of Pakistan through project No. 20-1484/R\&D/09 and the Abdus
Salam International Center for Theoretical Physics (ICTP) for support through
the Associateship Scheme.

\bigskip

\bigskip

\begin{figure}[h]
\centering
\subfigure[]{\label{fig:1(a)}\includegraphics[width=0.4\textwidth]{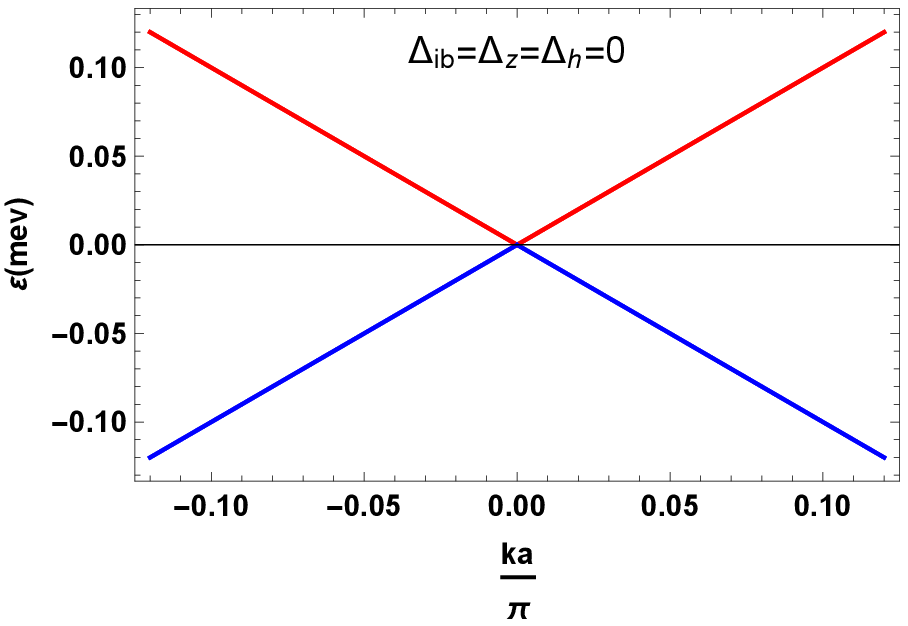}}
\subfigure[]{\label{fig:1(b)}\includegraphics[width=0.4\textwidth]{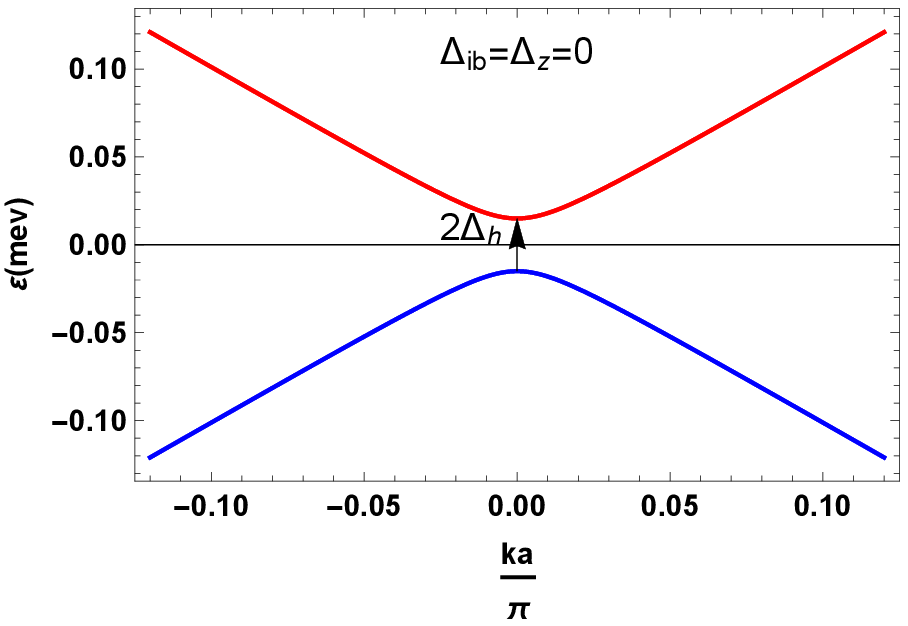}}
\subfigure[]{\label{fig:1(c)}\includegraphics[width=0.4\textwidth]{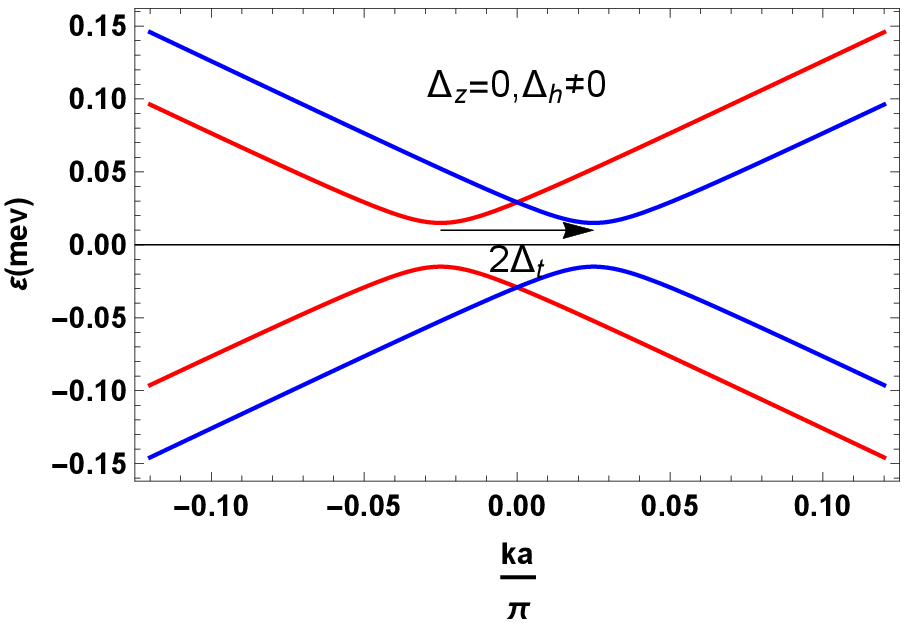}}
\subfigure[]{\label{fig:1(d)}\includegraphics[width=0.4\textwidth]{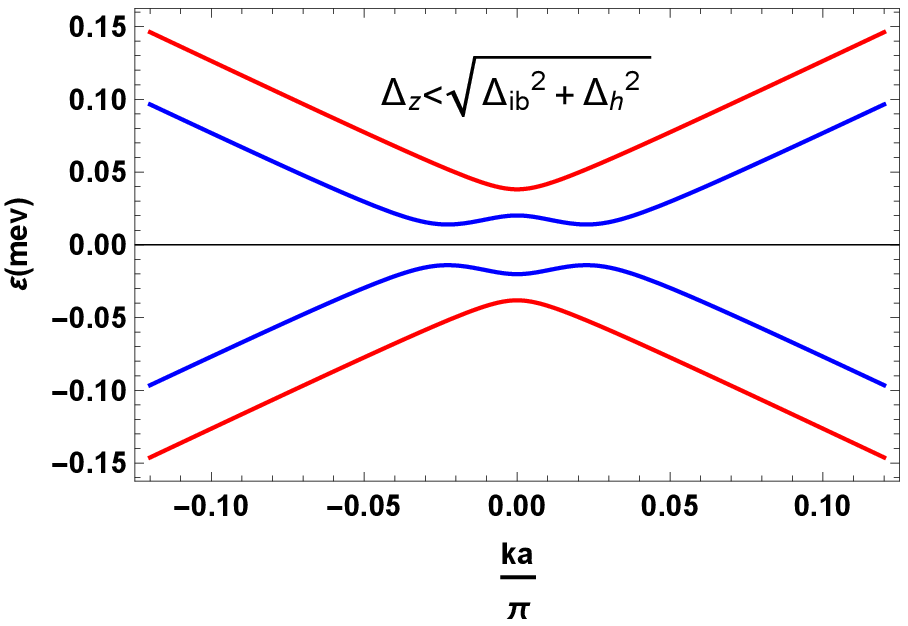}}
\subfigure[]{\label{fig:1(e)}\includegraphics[width=0.4\textwidth]{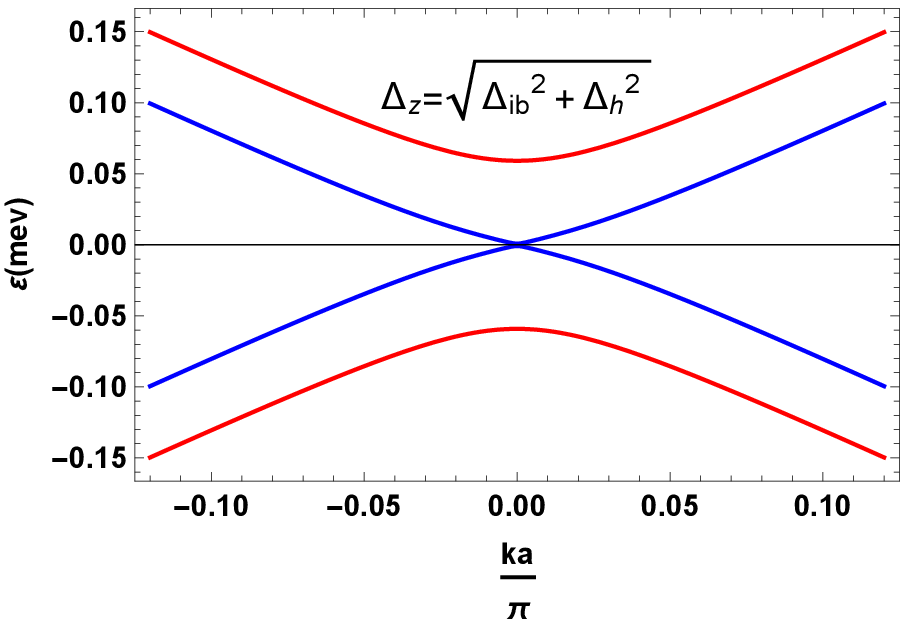}}
\subfigure[]{\label{fig:1(f)}\includegraphics[width=0.4\textwidth]{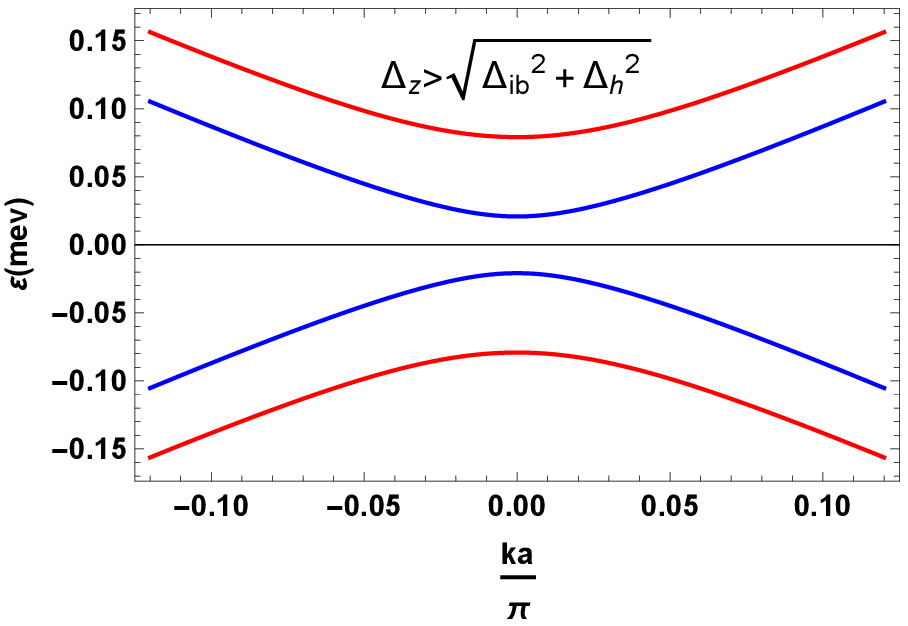}}
\caption{Schematic representation of surface band structure. (a) $\Delta
_{h}=\Delta_{ib}=\Delta_{z}=0$ (b) $\Delta_{h}\ne 0,$ $\Delta_{ib}%
=\Delta_{z}=0$ (c) $(\Delta_{h},\Delta_{ib}) \ne 0,$ $\Delta
_{z}=0$ (d)$\Delta_{z}<\sqrt{\Delta_{ib}^{2}+\Delta_{h}^{2}}$ (e) $\Delta_{z}=\sqrt{\Delta_{ib}^{2}+\Delta_{h}^{2}}$ (f) $\Delta_{z}>\sqrt{\Delta_{ib}^{2}+\Delta_{h}^{2}}$}%
\label{fig:1}
\end{figure}

\begin{figure}[h]
\centering
\subfigure[]{\label{fig:2(a)}\includegraphics[width=0.6\textwidth]{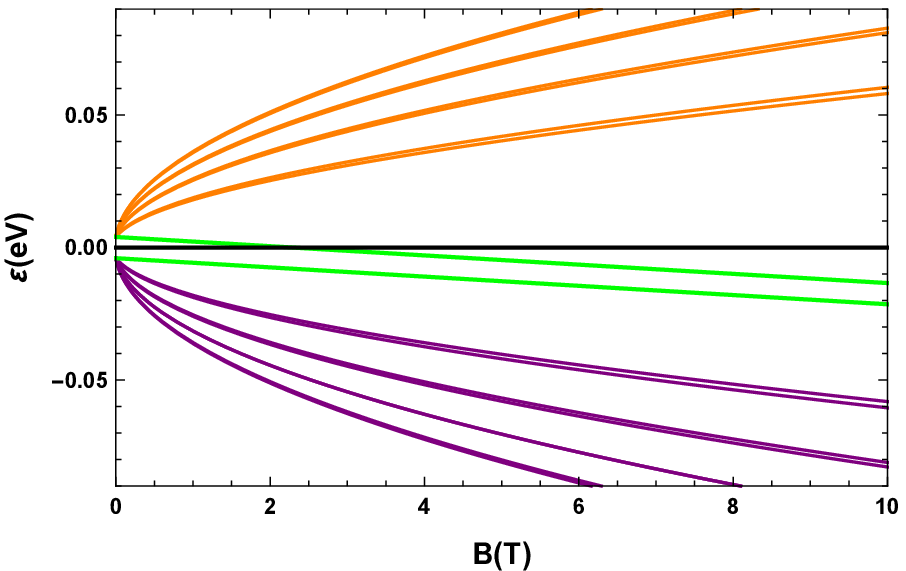}}
\subfigure[]{\label{fig:2(b)}\includegraphics[width=0.6\textwidth]{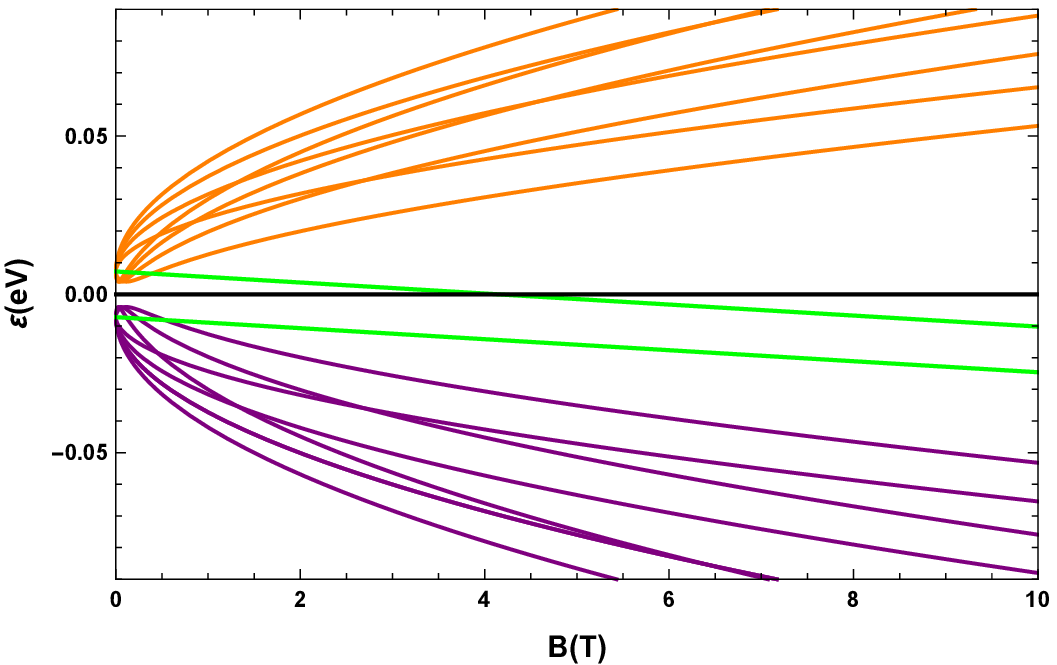}}

\caption{Landau level energies as a function of magnetic field (B) in units of
\ Tesla with hybridization energy $\Delta_{h}=0.004eV$, and
Zeeman energy $\Delta_{z}=0.0017BeVT^{-1}.$ a: $\Delta_{ib}=0$, b: $\Delta_{ib}=0.006eV$. The $n=0$ landau levels are represented by green lines, shifting from an electron-like and hole like set to both being hole-like as the magnetic field is increased. In 2(b) the crossing of LLs can be clearly seen as a result of finite $\Delta_{ib}$ }
\label{fig:2}
\end{figure}

\begin{figure}
[ptb]
\begin{center}
\includegraphics[
height=3.9721in,
width=3.378in
]%
{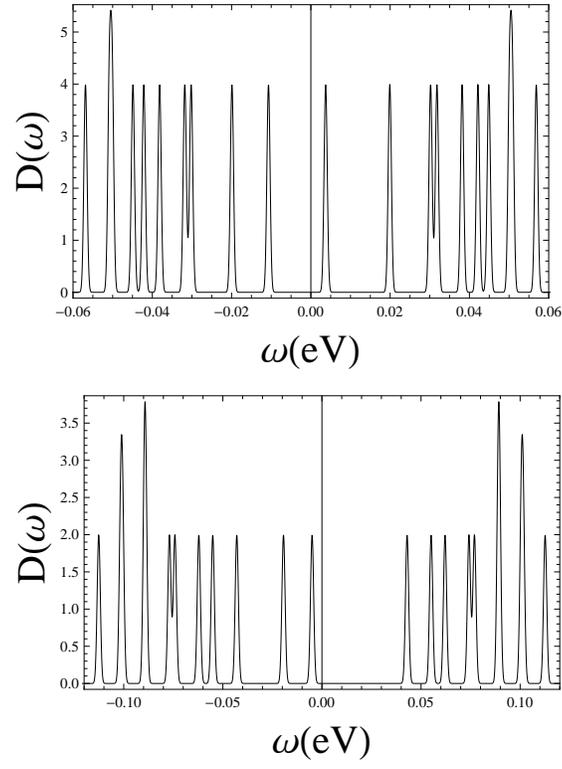}%
\caption{Density of states for thin film topological insulator in a magnetic
field in units of $eB/h$. Upper (NI) Density of states in normal insulator
phase $(\Delta_{z}<\sqrt{\Delta_{ib}^{2}+\Delta_{h}^{2}})$ . Lower Density of
states in quantum Hall phase phase$(\Delta_{z}>\sqrt{\Delta_{ib}^{2}%
+\Delta_{h}^{2}})$.}%
\end{center}
\end{figure}

\begin{figure}
[ptb]
\begin{center}
\includegraphics[
height=3.9704in,
width=5.3835in
]%
{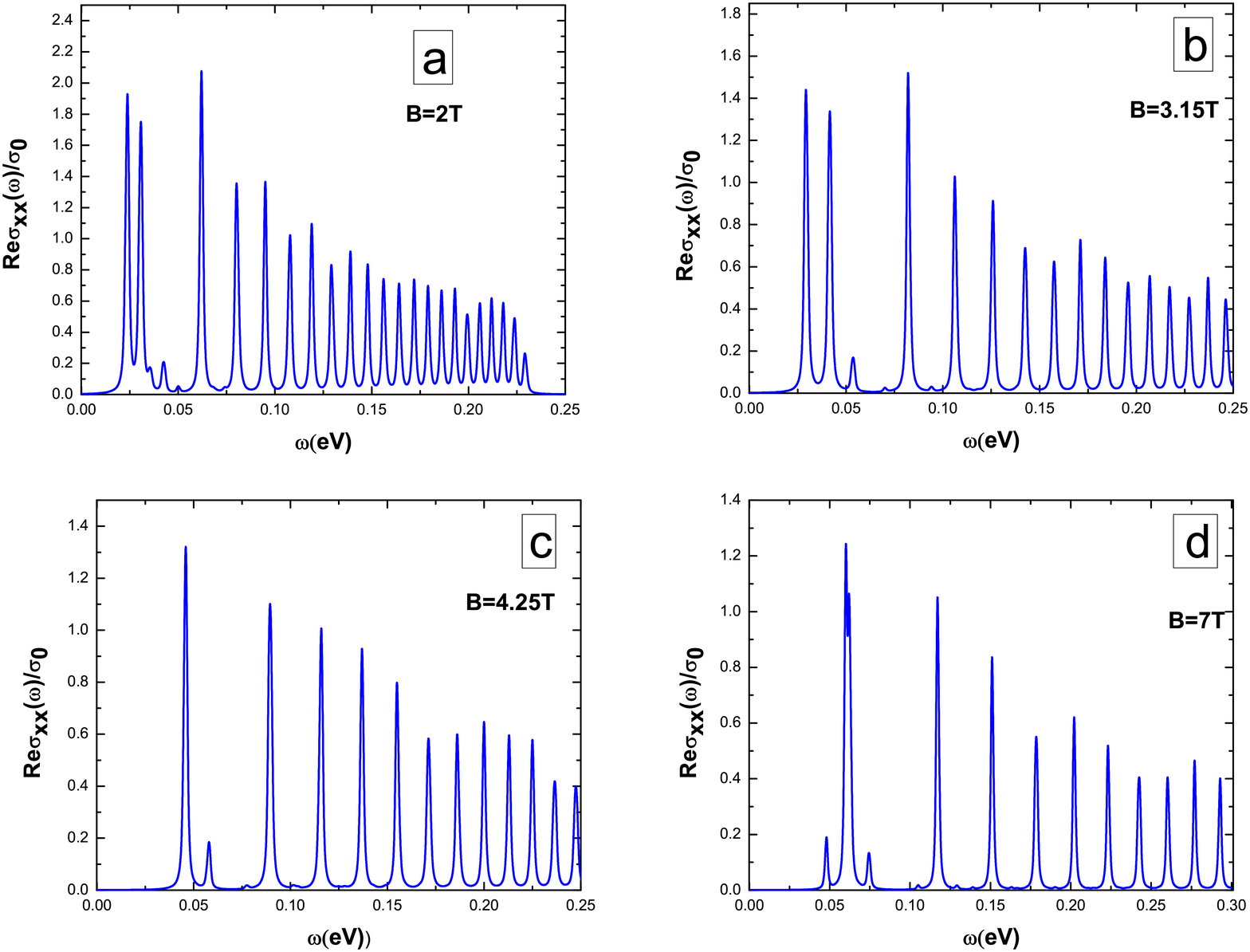}%
\caption{Real part of the longitudinal conductivity $\sigma_{xx}(\omega)$ of
thin film topological insulator in units of $e^{2}/\hbar$ as a function of
$\omega$ in $eV$. (a,b) normal insulator phase, (c) at CNP and (d) quantum
Hall phase. All these absorption peaks are resulted from the transition across
$\mu =0$.}%
\end{center}
\end{figure}
\begin{figure}[h]
\centering
\subfigure[]{\label{fig:5(a)}\includegraphics[width=0.4\textwidth]{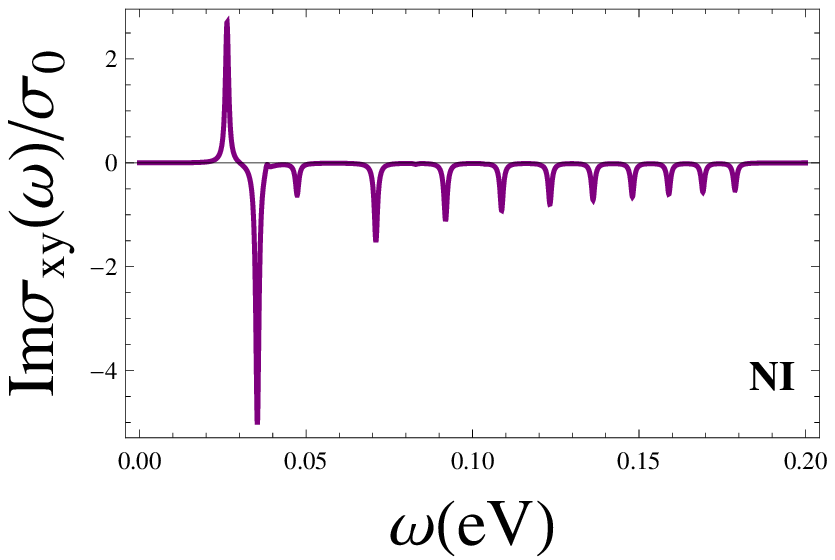}}
\subfigure[]{\label{fig:5(b)}\includegraphics[width=0.4\textwidth]{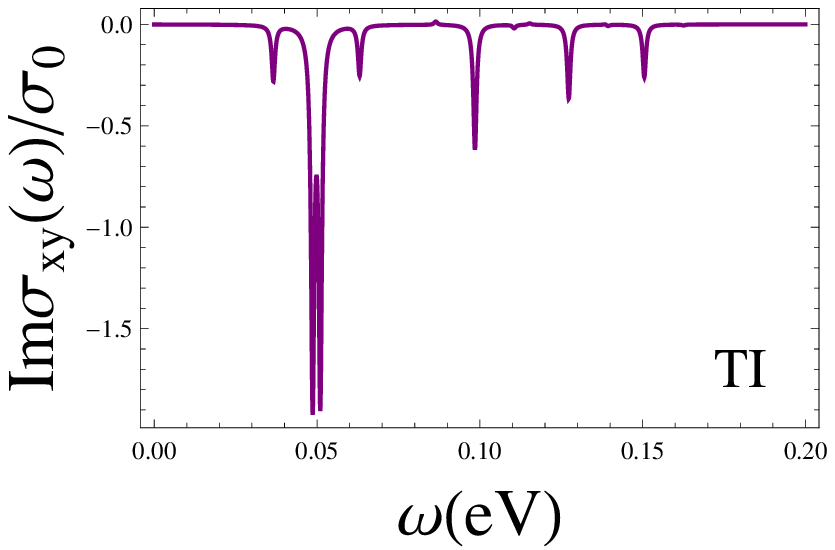}}
\caption{Imaginary part of the transverse conductivity $\sigma_{xy}(\omega)$of
thin film topological insulator in units of $e^{2}/\hbar$ as a function of
\ $\omega$ in $eV$. a: Normal phase, b: QH phase}
\label{fig:5}
\end{figure}

\begin{figure}
[ptb]
\begin{center}
\includegraphics[
height=3.007in,
width=4.369in
]%
{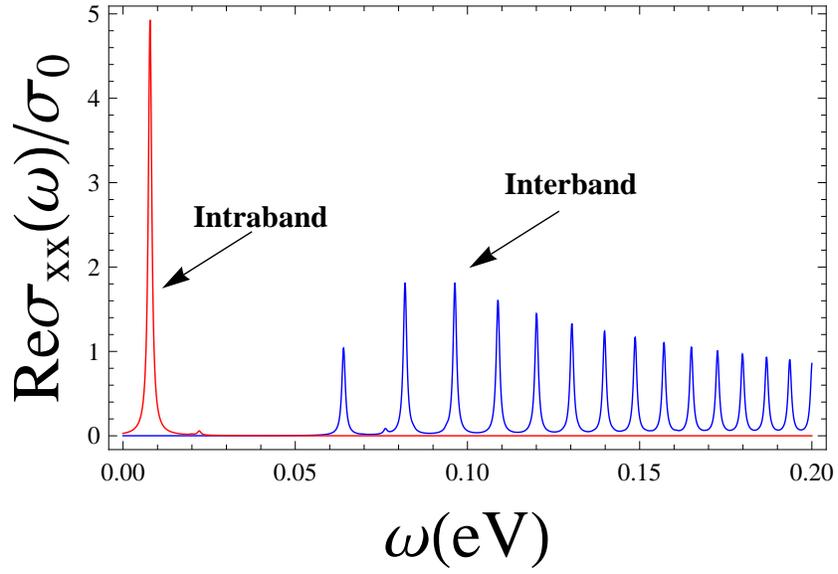}%
\caption{Real part of the longitudinal conductivity $\sigma_{xx}(\omega)$ of
thin film topological insulator in units of e$^{2}/\hbar$ as a function of
$\omega$ in $eV$ for $B=2$T. Red peak represents absorption peak for intraband
transition while blue peaks represent absorption peaks for interband
transitions.}%
\end{center}
\end{figure}
\begin{figure}
[ptb]
\begin{center}
\includegraphics[
height=3.4653in,
width=5.2857in
]%
{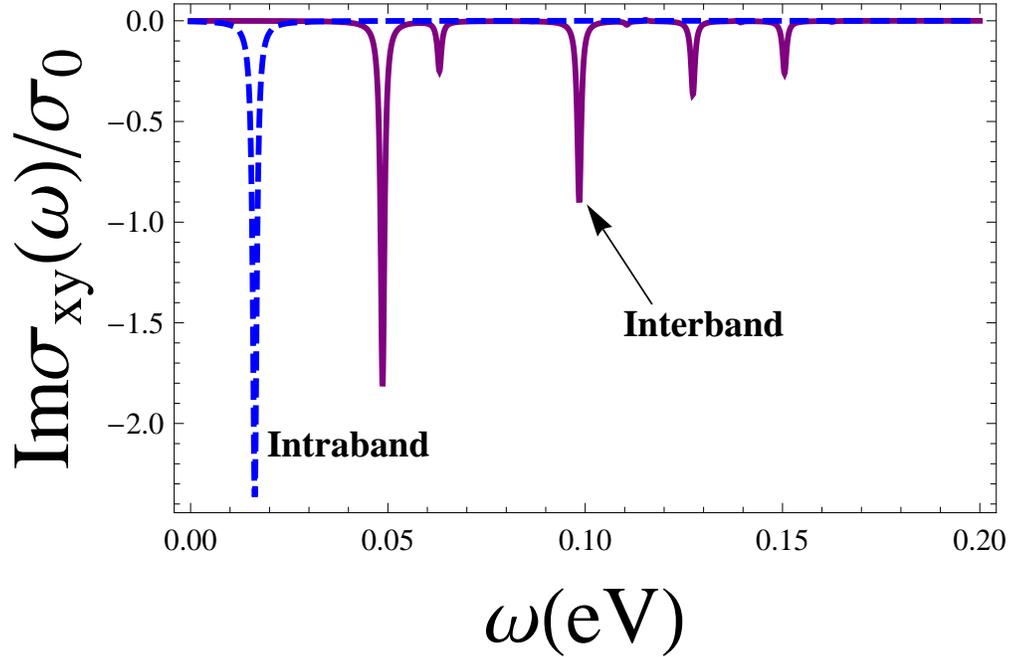}%
\caption{Imaginary part of the transverse conductivity $\sigma_{xy}(\omega)$of
thin film topological insulator in units of $e^{2}/\hbar$ as a function of
$\omega$ in $eV$ for $B=5$T. Blue dotted peak represents absorption peak for
intraband transition while purple peaks represent absorption peaks for
interband transitions.}%
\end{center}
\end{figure}

\begin{figure}
[ptb]
\begin{center}
\includegraphics[
height=3.9712in,
width=5.2269in
]%
{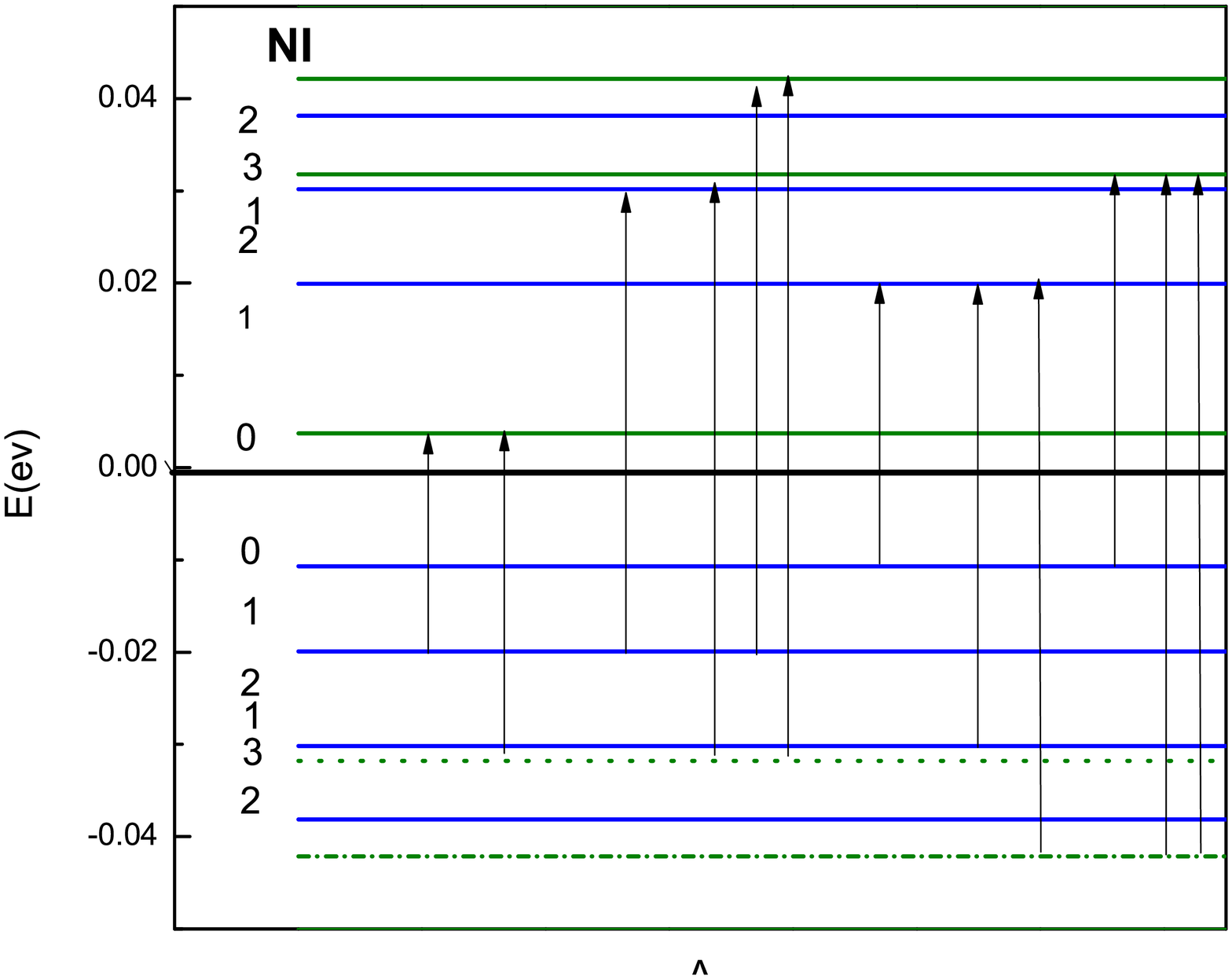}%
\caption{Schematic representation of the allowed transitions between Landau
levels in normal insulator phase at $B=2$T.}%
\end{center}
\end{figure}

\begin{figure}
[ptb]
\begin{center}
\includegraphics[
height=3.9712in,
width=5.2269in
]%
{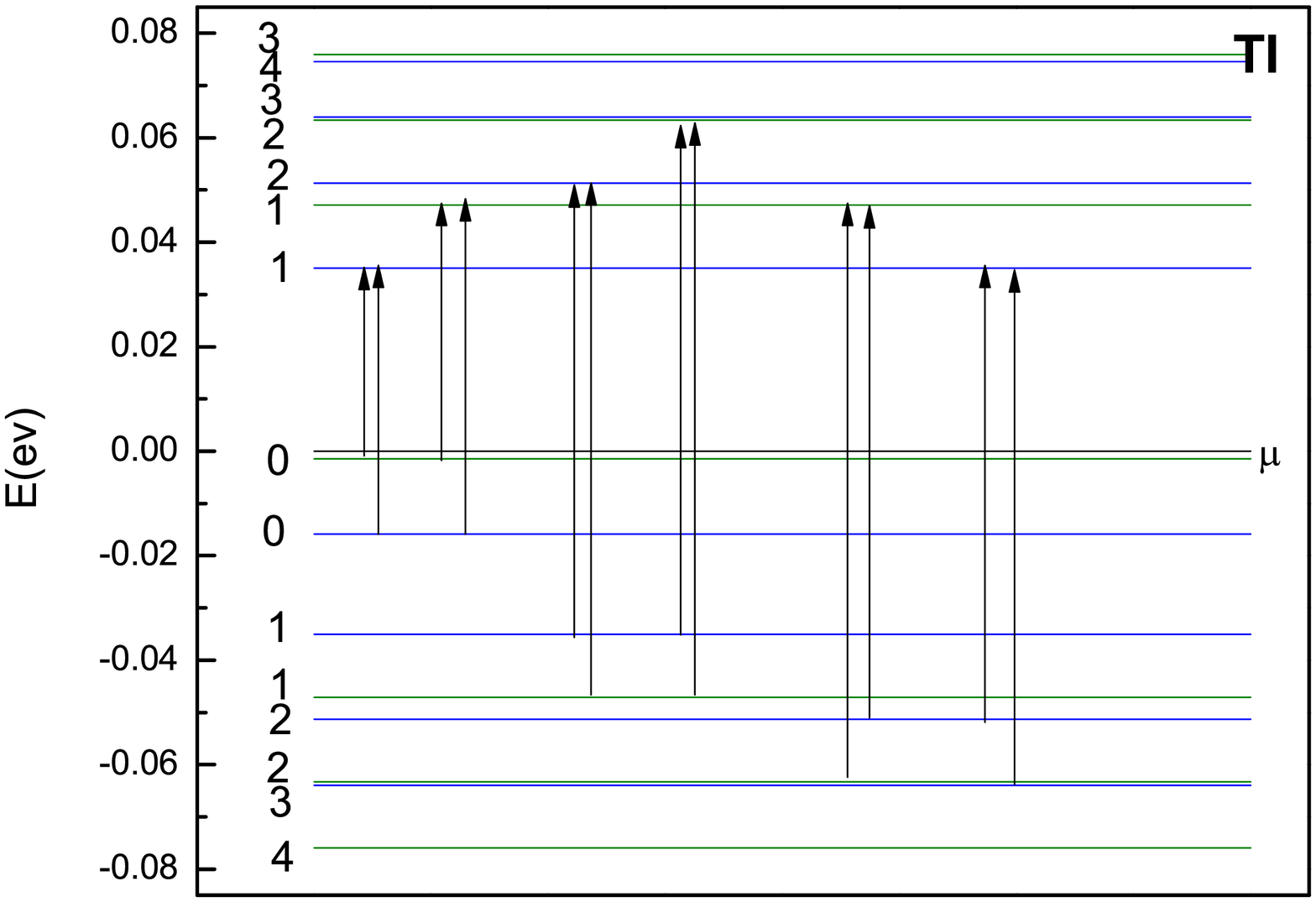}%
\caption{Schematic representation of the allowed transitions between Landau
levels in quantum Hall phase at $B=5$T.}%
\end{center}
\end{figure}

\begin{figure}[ptb]
\begin{center}
\includegraphics[
height=3.6781in,
width=5.6498in
]%
{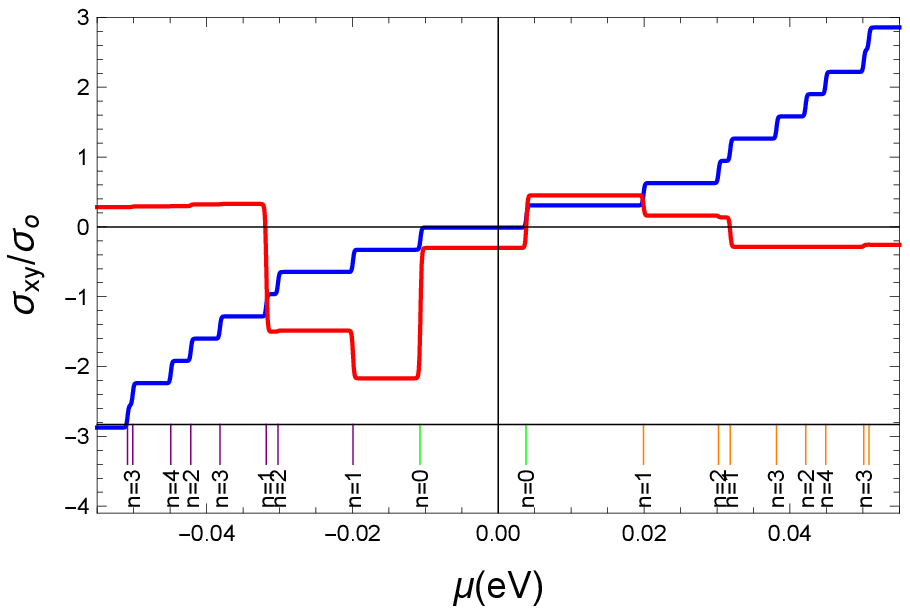}
\caption{Static Hall conductivity(blue) and optical Hall conductivity(red) as
a function of chemical potential at magnetic field $B=2$T. For optical Hall
conductivity $w=0.02eV.$}
\end{center}
\end{figure}

\begin{figure}
[ptb]
\begin{center}
\includegraphics[
height=3.6781in,
width=5.6498in
]%
{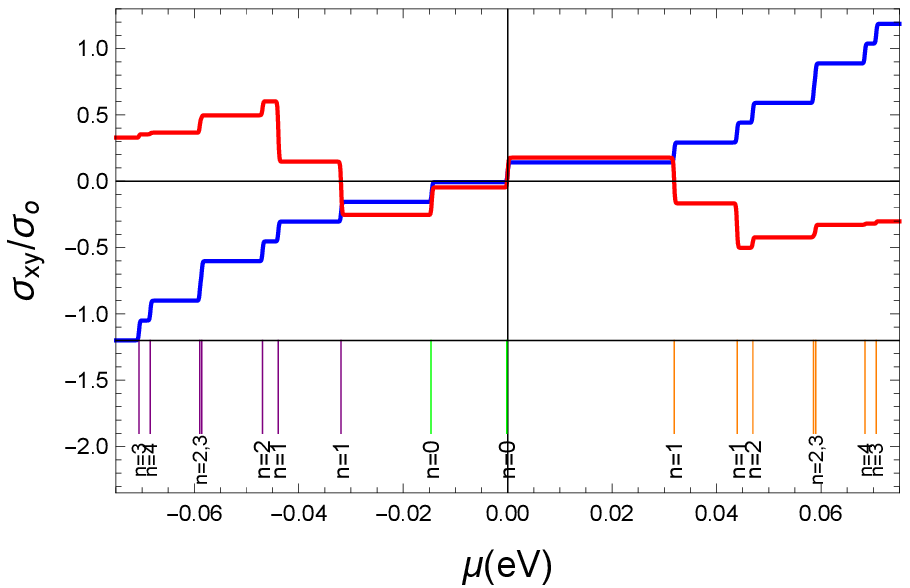}%
\caption{Static Hall conductivity(blue) and optical Hall conductivity(red) as
a function of chemical potential at magnetic field $B=4.25$T. For optical Hall
conductivity $w=0.02eV.$}%
\end{center}
\end{figure}

\begin{figure}
[ptb]
\begin{center}
\includegraphics[
height=3.6772in,
width=5.7545in
]%
{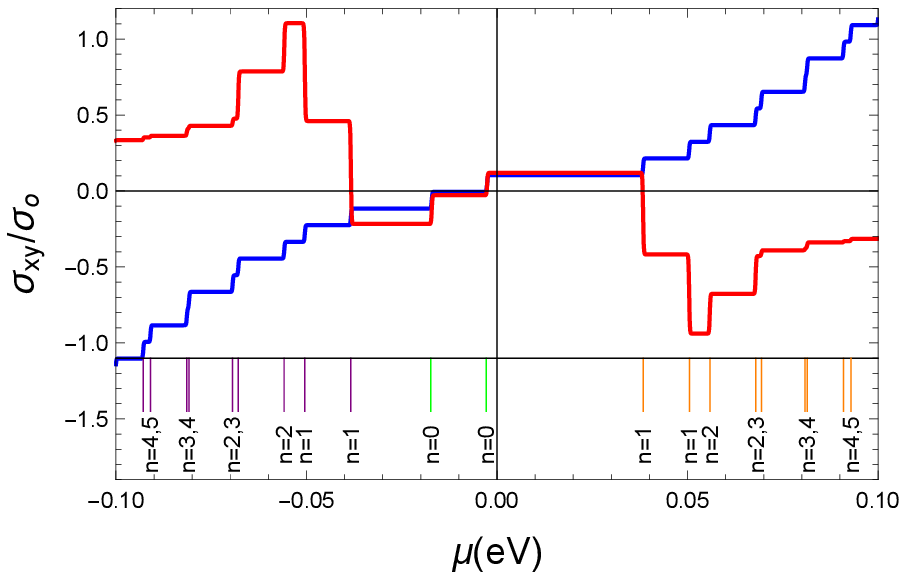}%
\caption{Static Hall conductivity(blue) and optical Hall conductivity(red) as
a function of chemical potential at magnetic field $B=5.8$T. For optical Hall
conductivity $w=0.02eV.$}%
\end{center}
\end{figure}

\begin{figure}[h]
\centering
\subfigure[]{\label{fig:13(a)}\includegraphics[width=0.7\textwidth]{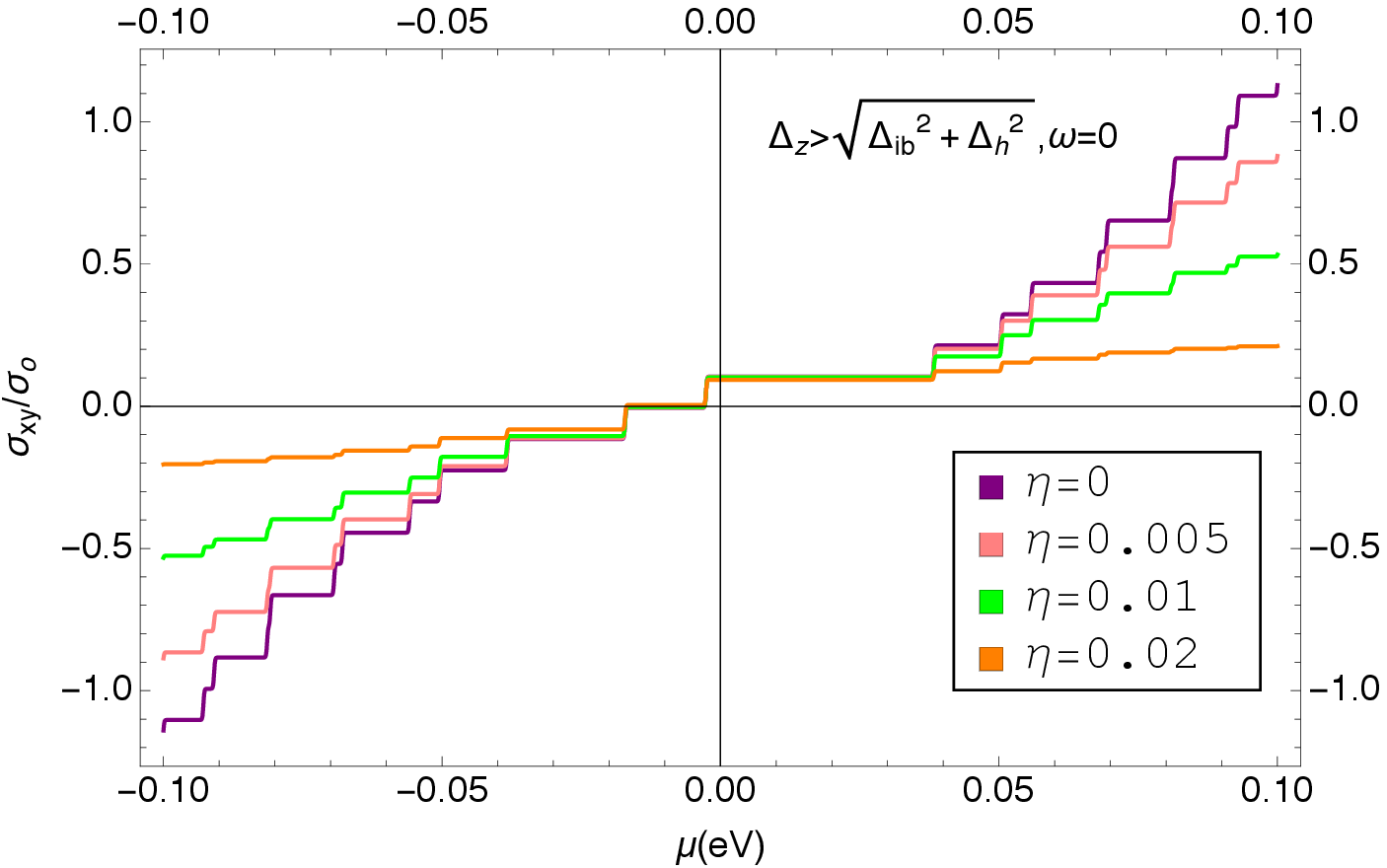}}
\subfigure[]{\label{fig:13(b)}\includegraphics[width=0.7\textwidth]{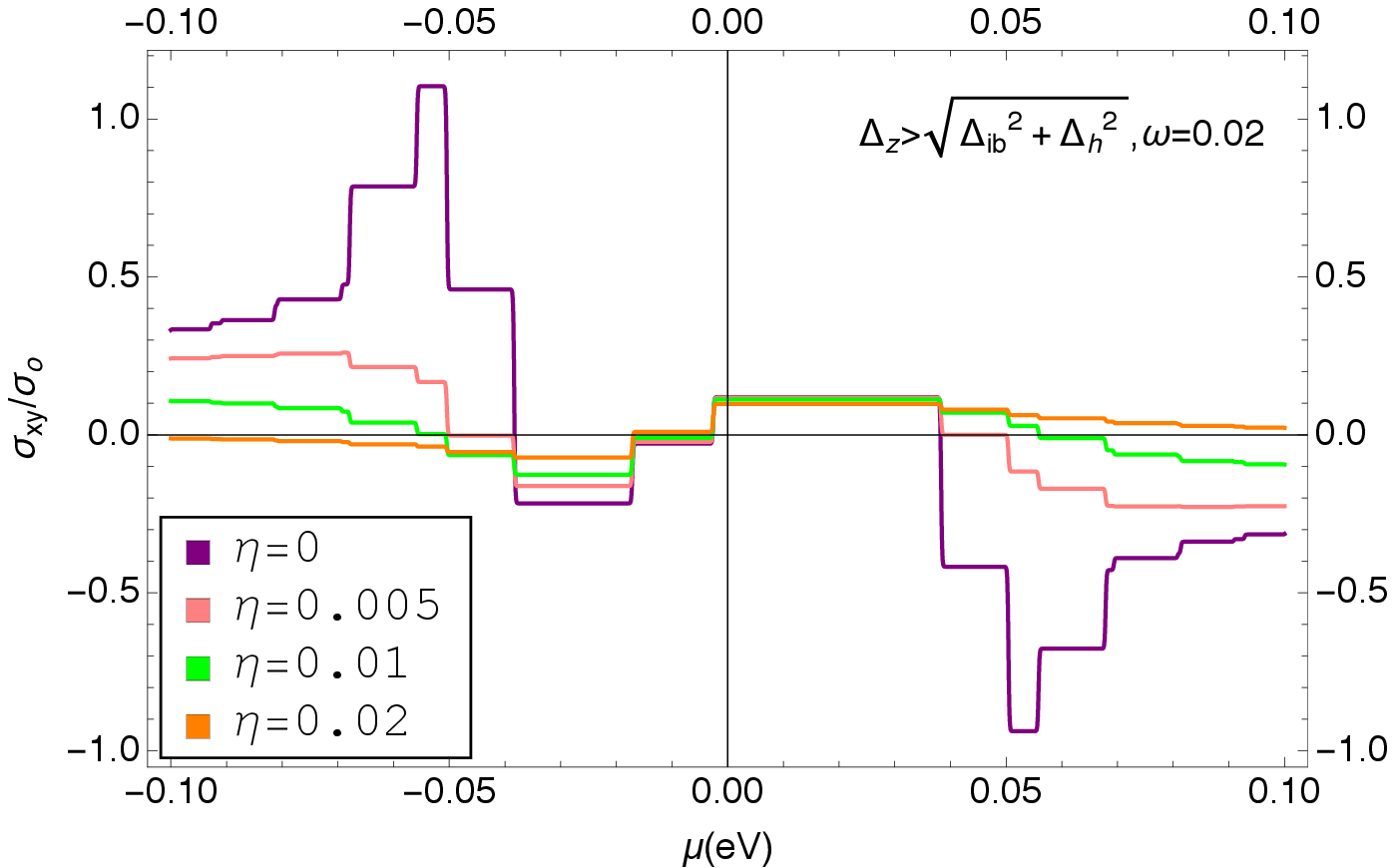}}
\caption{ (a)Dc Hall conductivity for different values of $\eta$ in Quantum Hall phaase; $\Delta_{z}>\sqrt{\Delta_{ib}^{2}+\Delta_{h}^{2}}$. The structure associated with $n=0$ LLs remains while the steps for large $|n|$ diminish as $\eta$ is increased. (b) For optical Hall
conductivity the $n=0$ LLs remains robust while for large $|n|$ the step-like structure diminishes more rapidly with increasing $\eta$ as compared to static case. Here $\omega$ is set at $0.02eV.$}
\label{fig:13}
\end{figure}

\begin{figure}[h]
\centering
\subfigure[]{\label{fig:14(a)}\includegraphics[width=0.7\textwidth]{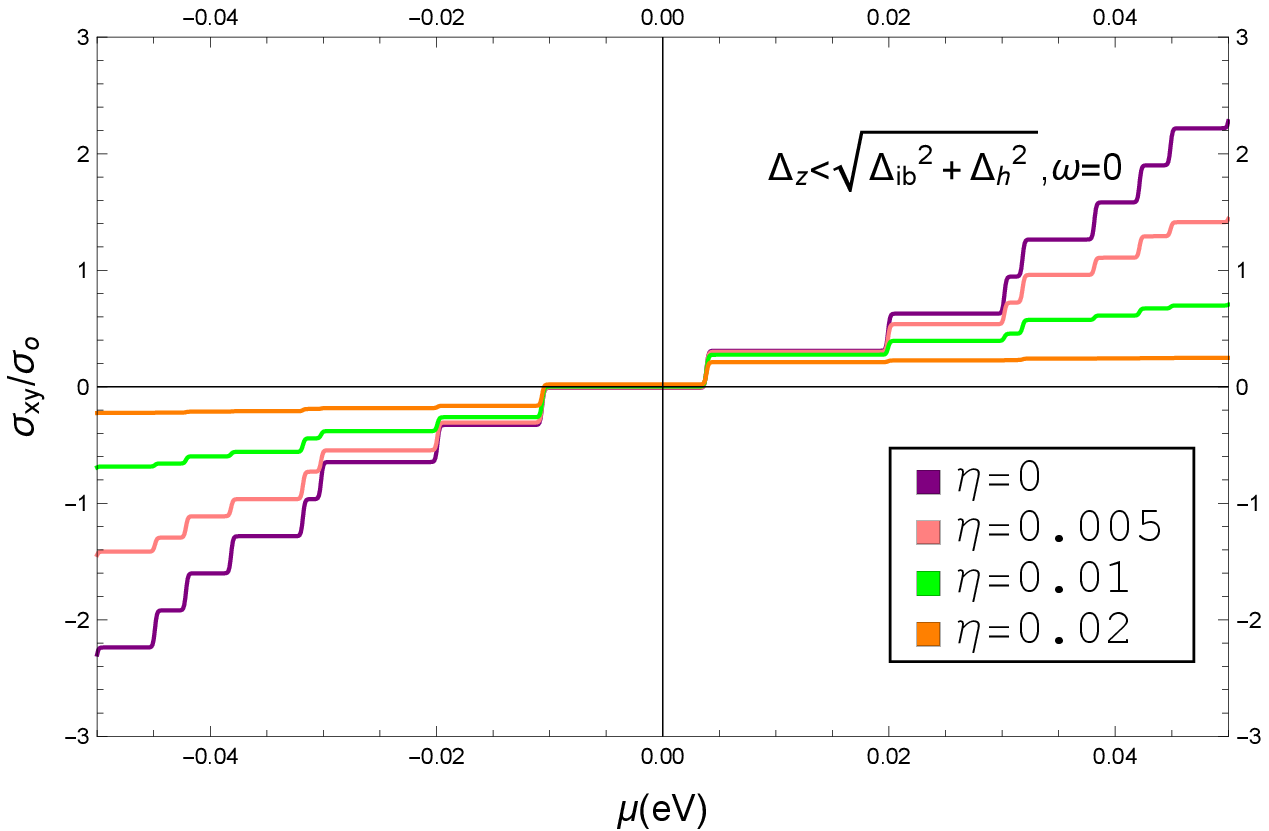}}
\subfigure[]{\label{fig:14(b)}\includegraphics[width=0.7\textwidth]{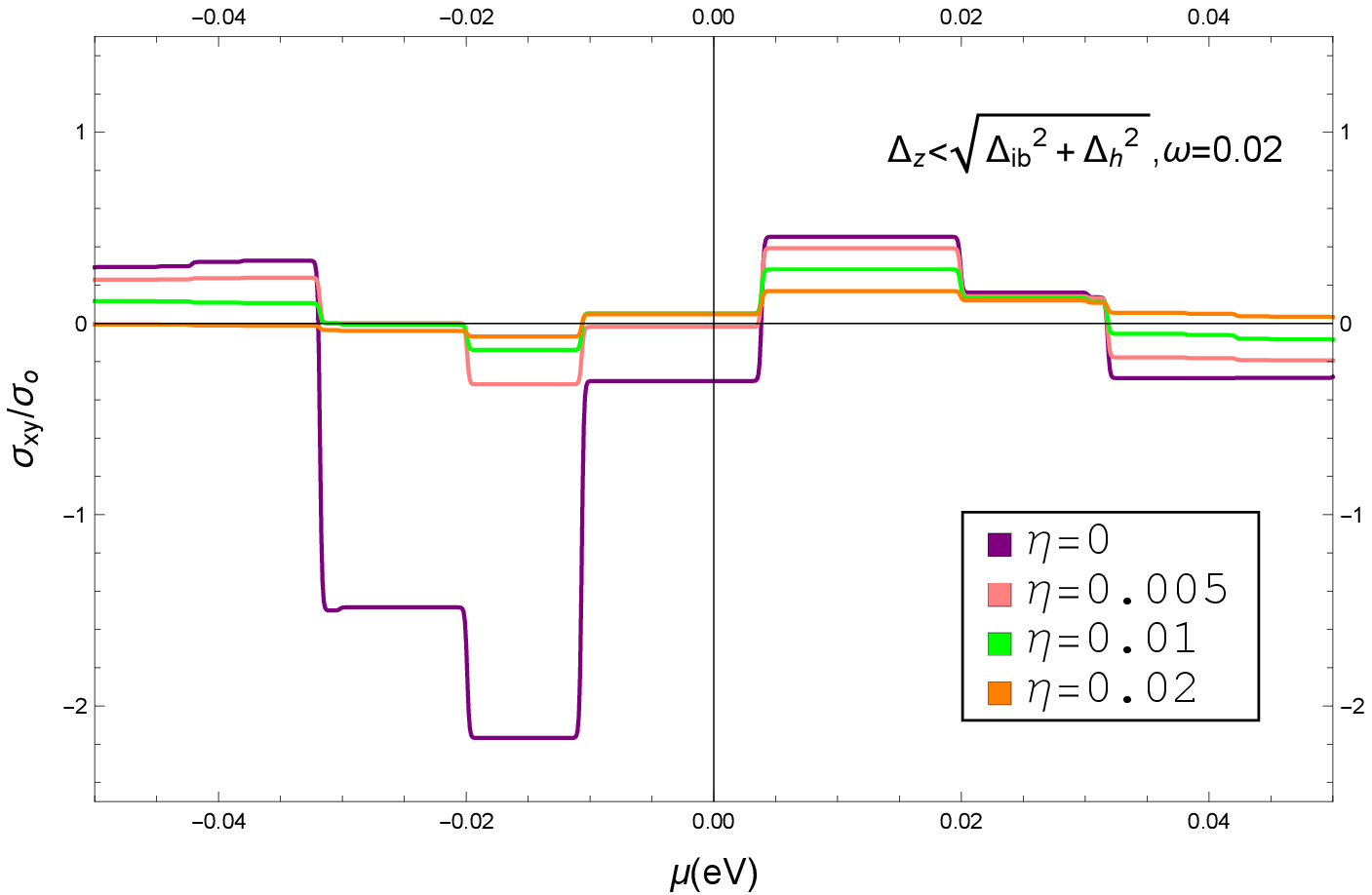}}
\caption{(a)Dc Hall conductivity for different values of $\eta$ in NI phaase; $\Delta_{z}<\sqrt{\Delta_{ib}^{2}+\Delta_{h}^{2}}$. The structure associated with $n=0$ LLs remains while the steps for large $|n|$ diminish as $\eta$ is increased. (b) For optical Hall
conductivity the $n=0$ LLs remains robust while for large $|n|$ the step-like structure diminishes more rapidly with increasing $\eta$ as compared to static case. Here $\omega$ is set at $0.02eV.$}
\label{fig:14}
\end{figure}

\begin{figure}
[ptb]
\begin{center}
\includegraphics[
height=2.6333in,
width=4.0983in
]%
{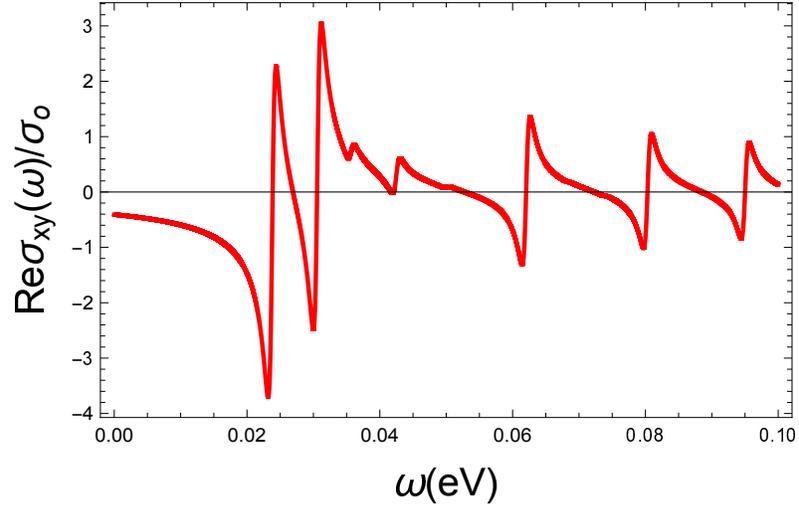}%
\caption{Optical Hall conductivity with $(\Delta
_{z}<\sqrt{\Delta_{ib}^{2}+\Delta_{h}^{2}})$ for chemical potential $\mu=0.$}%
\end{center}
\end{figure}

\begin{figure}
[ptb]
\begin{center}
\includegraphics[
height=2.7553in,
width=4.2333in
]%
{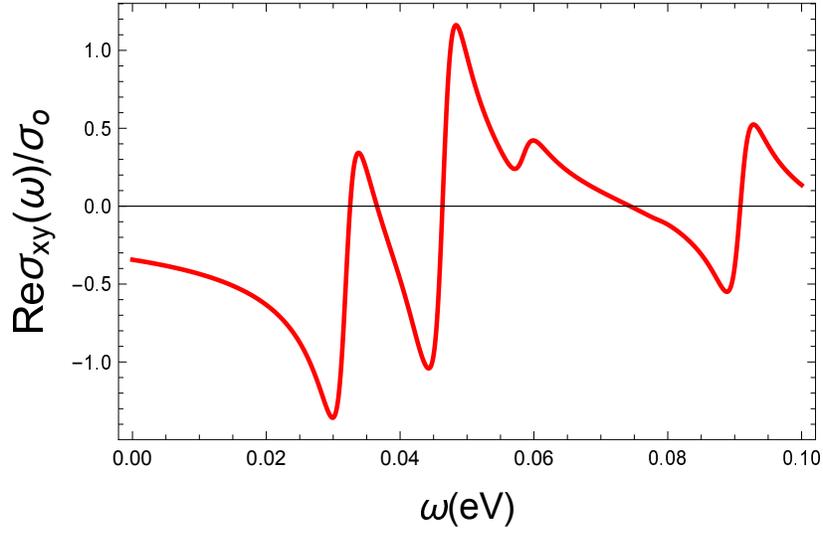}%
\caption{Optical Hall conductivity with $(\Delta
_{z}>\sqrt{\Delta_{ib}^{2}+\Delta_{h}^{2}})$ for chemical potential $\mu=0$
$(i.e.$optical Hall conductivity in quantum Hall phase).}%
\end{center}
\end{figure}

\end{document}